\documentclass[a4,center,fleqn]{NAR}

\copyrightyear{xx}
\pubdate{xx xx xx}
\pubyear{xx}
\jvolume{x}
\jissue{x}

\usepackage{times}

\usepackage{array}

\usepackage{booktabs}
\usepackage{balance}
\usepackage{arydshln}

\usepackage{placeins}

\usepackage[bottom]{footmisc}

\usepackage{enumitem}
\usepackage{ifthen}

\newenvironment{sciabstract}{%
\begin{quote} \bf}
{\end{quote}}

\usepackage{xspace}
\usepackage{amssymb,amsmath,epsfig}
\usepackage{bbding} 
\usepackage{algorithmicx,algorithm}
\usepackage[noend]{algpseudocode}

\algrenewcommand\algorithmicindent{0.8em}%

\algnewcommand\algorithmicforeach{\textbf{for each}}
\algdef{S}[FOR]{ForEach}[1]{\algorithmicforeach\ #1\ \algorithmicdo}

\usepackage[table]{xcolor} 
\usepackage{booktabs}
\usepackage{url}
\usepackage{float} 
\usepackage{longtable}
\usepackage{setspace}
\definecolor{myblue}{RGB}{245,245,245}
\usepackage[singlelinecheck=off]{caption}
\usepackage{lineno}

\usepackage[T1]{fontenc} 
\usepackage{array}

\usepackage{xparse}

\usepackage{tikz}
\usetikzlibrary{matrix, positioning, arrows.meta, calc, shapes, decorations,
  backgrounds, arrows, decorations.pathreplacing, automata}
\usepackage{pgfplots, siunitx}
\usepackage{hhline}
\usepackage{lmodern} 
\usepackage{tikz-qtree} 
\usepackage{forest}
\usetikzlibrary{patterns}

\usepackage{rotating}

\usepackage{array, longtable, makecell}
\usepackage{multirow}
\usepackage{textcomp}

\usepackage{mfirstuc}
\usepackage{titlecaps}

\usepackage{amsmath}

\usepackage{MnSymbol}

\usepackage{csvsimple}
\usepackage{pgf-pie}

\usepackage{pgfplots, pgfplotstable}

\usepackage{hyperref}
\usepackage{soul}

\newcommand{\nts}{\ensuremath{\text{\it nt}}\xspace}

\newcommand{\notes}[1]{}

\newcommand{\ith}[1]{\ensuremath{i^{{th}}}}

\newcount\permx
\newcount\permy
\def\permdot#1#2{
\permx=#1 \advance\permx by-1
\permy=#2 \advance\permy by-1
\psframe[fillcolor=black, fillstyle=solid]
(\permx,\permy)(#1, #2)
}

\newcommand{\score}{\ensuremath{\textit{score}}\xspace}

\newcommand{\vecx}{\ensuremath{\boldsymbol{x}}\xspace}
\newcommand{\vecy}{\ensuremath{\boldsymbol{y}}\xspace}

\newcommand{\vecystar}{\ensuremath{\vecy^*}\xspace}

\newcommand{\smallnt}[1]{\ensuremath{_{\mbox{\tiny PP}}}\xspace}

\iffalse

\else

\fi

\newcommand{\smallurl}[1]{{\scriptsize \url{#1}}}

\newcommand{\linearfold}{{LinearFold}\xspace}
\newcommand{\linearpartition}{{LinearPartition}\xspace}

\newcommand{\rnaalifold}{{RNAalifold}\xspace}
\newcommand{\linearalifold}{{LinearAlifold}\xspace}
\newcommand{\linalifold}{{LinAliFold}\xspace}
\newcommand{\rnastralign}{{RNAstralign}\xspace}

\newcommand{\nucA}{\ensuremath{\text{\sc a}}}
\newcommand{\nucU}{\ensuremath{\text{\sc u}}}
\newcommand{\nucC}{\ensuremath{\text{\sc c}}}
\newcommand{\nucG}{\ensuremath{\text{\sc g}}}

\newcommand{\panel}[1]{\large \sf {#1}}

\definecolor{intnull}{RGB}{213,229,255}
\definecolor{inteins}{RGB}{128,179,255}
\definecolor{intvier}{RGB}{42,127,255}
\definecolor{intdrei}{RGB}{0,85,212}
\definecolor{intvier}{RGB}{0,51,128}
\definecolor{intfunf}{RGB}{0,34,85}

\newcommand{\sarscovtwo}{{SARS-CoV-2}\xspace}
\newcommand{\sarscov}{{SARS-CoV}\xspace}
\newcommand{\sarsr}{{SARS-related}\xspace}
\newcommand{\linearturbofold}{{LinearTurboFold}\xspace}

\newsavebox\CBox

\def\namecite{\citet}

\newcommand{\pairables}{\ensuremath{\mathcal{P}}\xspace}

\newcommand{\E}{\operatornamewithlimits{\mathbb{E}}}

\begin{document} 

\title{\ \!\!LinearAlifold: Linear-Time Consensus Structure Prediction for RNA Alignments}


\author{\small
Apoorv Malik$^{\dagger,\bullet}$ 
Liang Zhang$^{\dagger,\bullet}$
Milan Gautam$^\dagger$
Ning Dai$^\dagger$  
Sizhen Li$^\dagger$
He Zhang$^\dagger$
David H. Mathews$^{\diamond,\circ,\clubsuit}$
Liang Huang$^{\dagger,\ddagger,\bullet,\ast}$}
 
\address{
 $^{\dagger}$School of EECS and $^\ddagger$Dept.~of Biochemistry \& Biophysics,
Oregon State University, Corvallis, OR 97330, USA,
$^{\diamond}$Dept.~of Biochemistry \& Biophysics, 
$^{\circ}$Center for RNA Biology, and
$^{\clubsuit}$Dept.~of Biostatistics \& Computational Biology,
University of Rochester Medical Center, Rochester, NY 14642, USA,
$^\bullet$Equal contribution.
{$^\ast$Corresponding author: {\scriptsize\tt liang.huang.sh@gmail.com}}
}






\maketitle
%

\marginparwidth2cm


\vspace{-0.5cm}
\begin{sciabstract}
Predicting the consensus structure of a set of aligned RNA homologs
is a convenient method to find conserved structures
in an RNA genome,
which has many applications including 
viral diagnostics and therapeutics.
However, the 
most commonly used tool
for this task, 
\rnaalifold, 
 is prohibitively slow for long sequences, due to a cubic scaling with the sequence length,
taking over a day 
on 400 \sarscovtwo and \sarsr genomes ($\sim$30,000\nts). 
We present \linearalifold, 
a much faster alternative 
that scales linearly with both the sequence length 
and the number of sequences,
based on our work \linearfold that folds a single RNA in linear time.  
Our work is orders of magnitude faster than \rnaalifold (0.7 hours on the above 400 genomes, or $\sim36\times$ speedup)
and achieves higher accuracies when compared to a database of known structures.
More interestingly, \linearalifold's prediction on \sarscovtwo correlates well with experimentally determined structures,
substantially outperforming \rnaalifold.
Finally, \linearalifold supports two energy models (Vienna and BL*) and four  modes: minimum free energy (MFE), maximum expected accuracy (MEA), ThreshKnot, 
and stochastic sampling, 
each of which takes under an hour for hundreds of \sarscov variants.
Our resource is at:\\ 
{\small \url{https://github.com/LinearFold/LinearAlifold}} (code) and
{\small \url{http://linearfold.org/linear-alifold}} (server). 
\end{sciabstract}


\section*{Introduction}
\label{sec:intro}


Ribonucleic acids (RNA) are involved in many cellular processes~\cite{eddy:2001,Doudna+cech:2002,bachellerie+:2002},
and most of RNA secondary structures are highly conserved across evolution to maintain their functionalities in spite of changes to the sequence~\cite{nawrocki+:2013,brown+:1992,ritz+:2013}.
Thus, predicting the {\em consensus structure} for a set of aligned RNA homologs is more accurate than predicting the structure for a single sequence and it is
useful for identifying conserved regions, which can be used for diagnostics and therapeutics.
For this task, \rnaalifold~\cite{hofacker+:2002,bernhart+:2008} is a widely used tool 
to predict consensus structures for aligned RNA homologs
that considers both thermodynamic stability and sequence covariation.
However, its cubic runtime (against sequence length $n$) makes it difficult to be applied to long sequences
such as \sarscovtwo genomes ($n\simeq 30,000 \nts$),
requiring over a day for 400 such genomes.
As an alternative, \linearturbofold \cite{li+:2021} is an iterative fold-and-align tool (thus does not need alignment as input) that scales linearly with sequence length, but quadratically with the number of sequences ($k$).
This limits its use case to only about 30 \sarscovtwo variants
while 
it is often helpful 
to include hundreds of such 
genomes to account for as much sequence variation as possible, as new variants emerge rapidly.
So there is a critical need to develop a fast consensus folding tool that scales linearly
with {\em both} $n$ and $k$.
On the other hand, beyond predicting minimum free energy (MFE) consensus structures, it is also useful 
to calculate the {\em consensus partition function} and {\em consensus base-pairing probabilities} (BPPs), which are widely used in many downstream tasks such as maximum expected accuracy (MEA) folding~\cite{Knudsen+Hein:2003,do+:2006,lu+:2009},
ThreshKnot~\cite{zhang+:2019},
and stochastic sampling from the ensemble~\cite{ding+lawrence:2003,zhang+:2020_LS}. 
However, \rnaalifold's partition function mode is even slower than its MFE mode (often by $10\times$ or more), and 
both its partition function and stochastic sampling modes fail to run on \sarscovtwo  
(for any $k>1$) due to overflow. 

\begin{figure*}[!t]
\centering
\begin{tabular}{cc} 

\raisebox{0pt}[\height][0pt]{\panel{A}} & \raisebox{0pt}[\height][0pt]{\panel{B} {\scriptsize ($k=30$)}} \\
\multirow{2}{*}[4.2cm]{\includegraphics[width=0.48\linewidth]{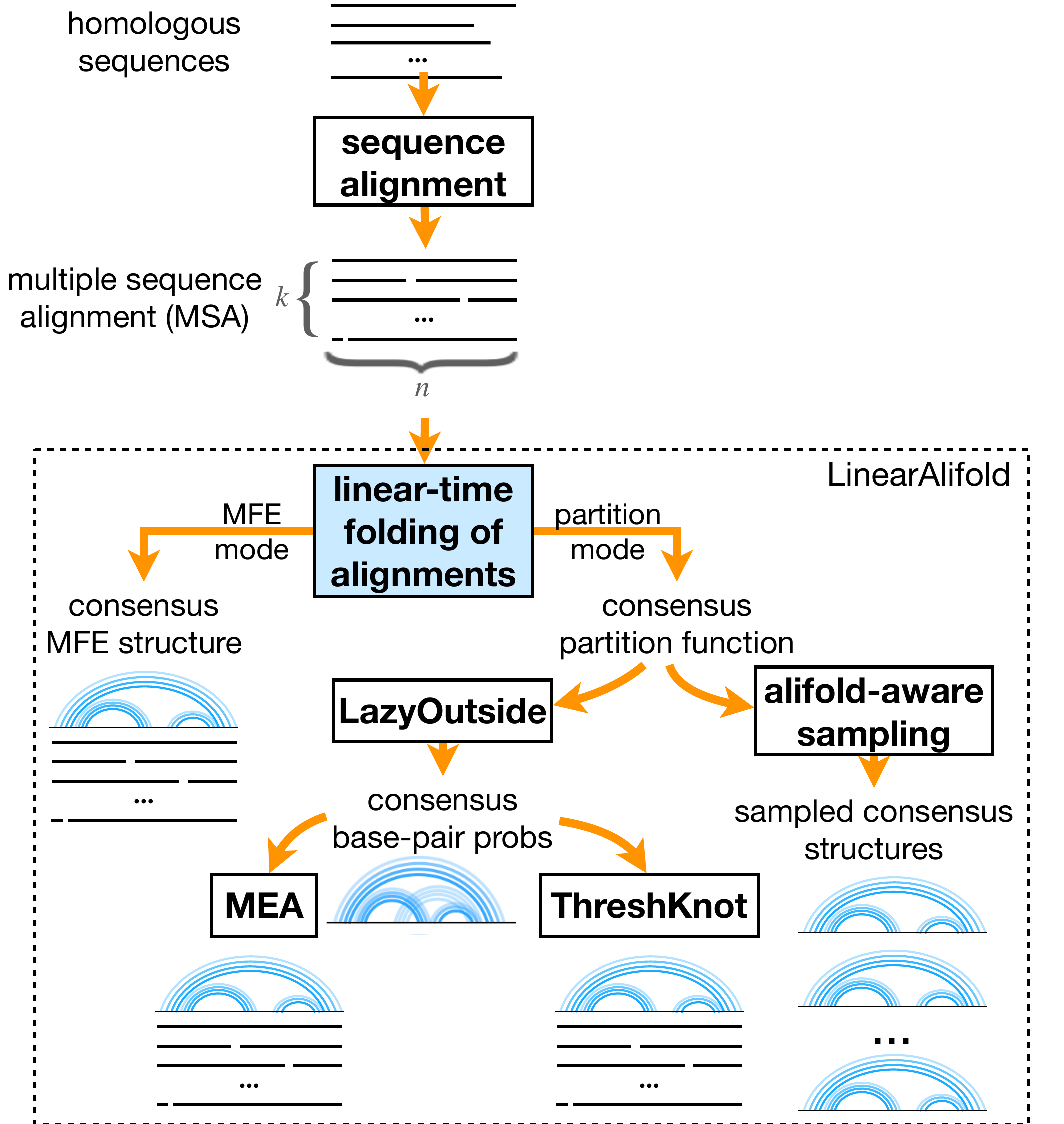}} & 
\includegraphics[width=0.47\linewidth]{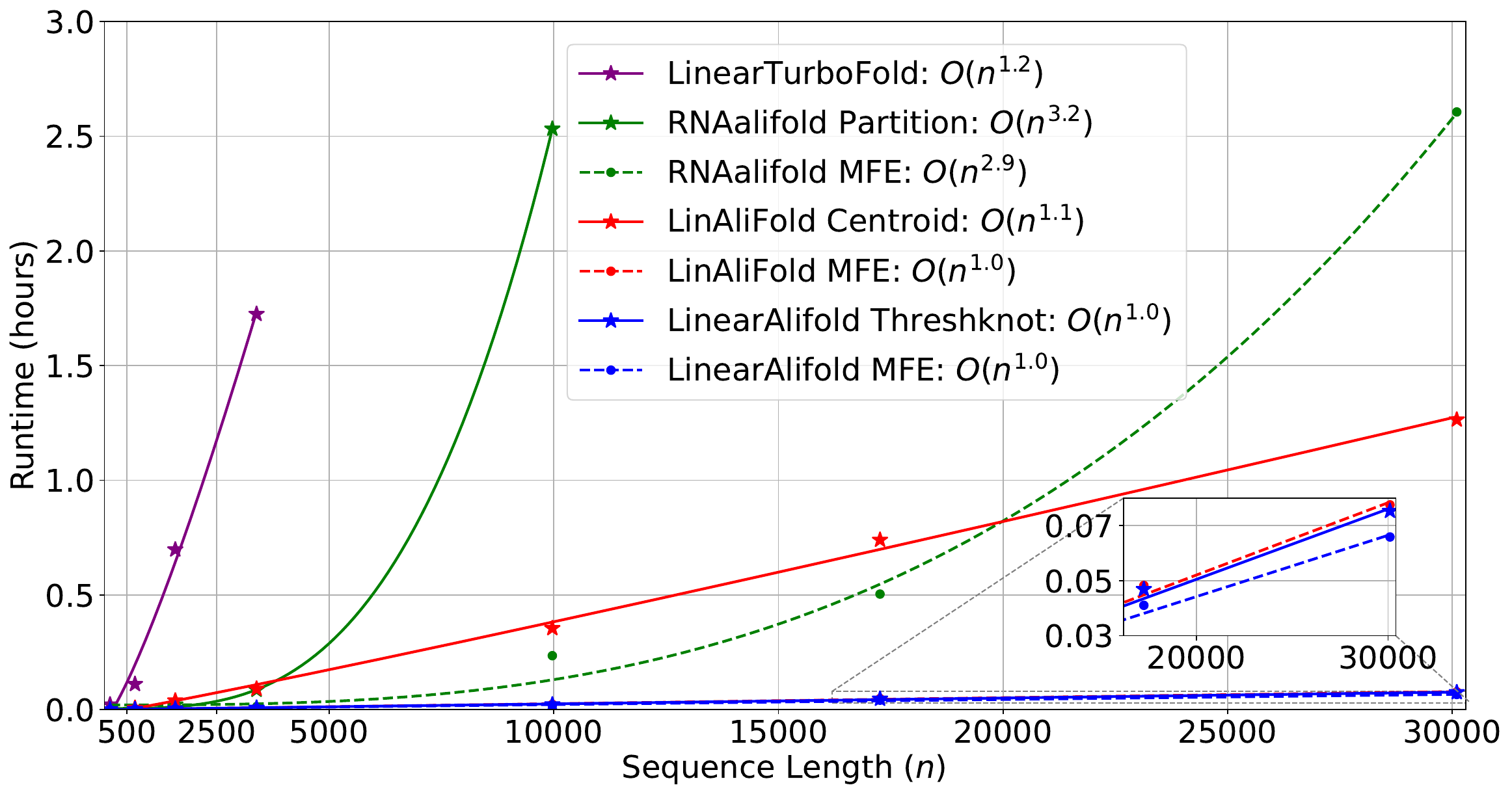} \\ 

& \raisebox{0pt}[\height][0pt]{\panel{C} {\scriptsize ($n\simeq 30,000\nts$)}} \\
& \includegraphics[width=0.47\linewidth]{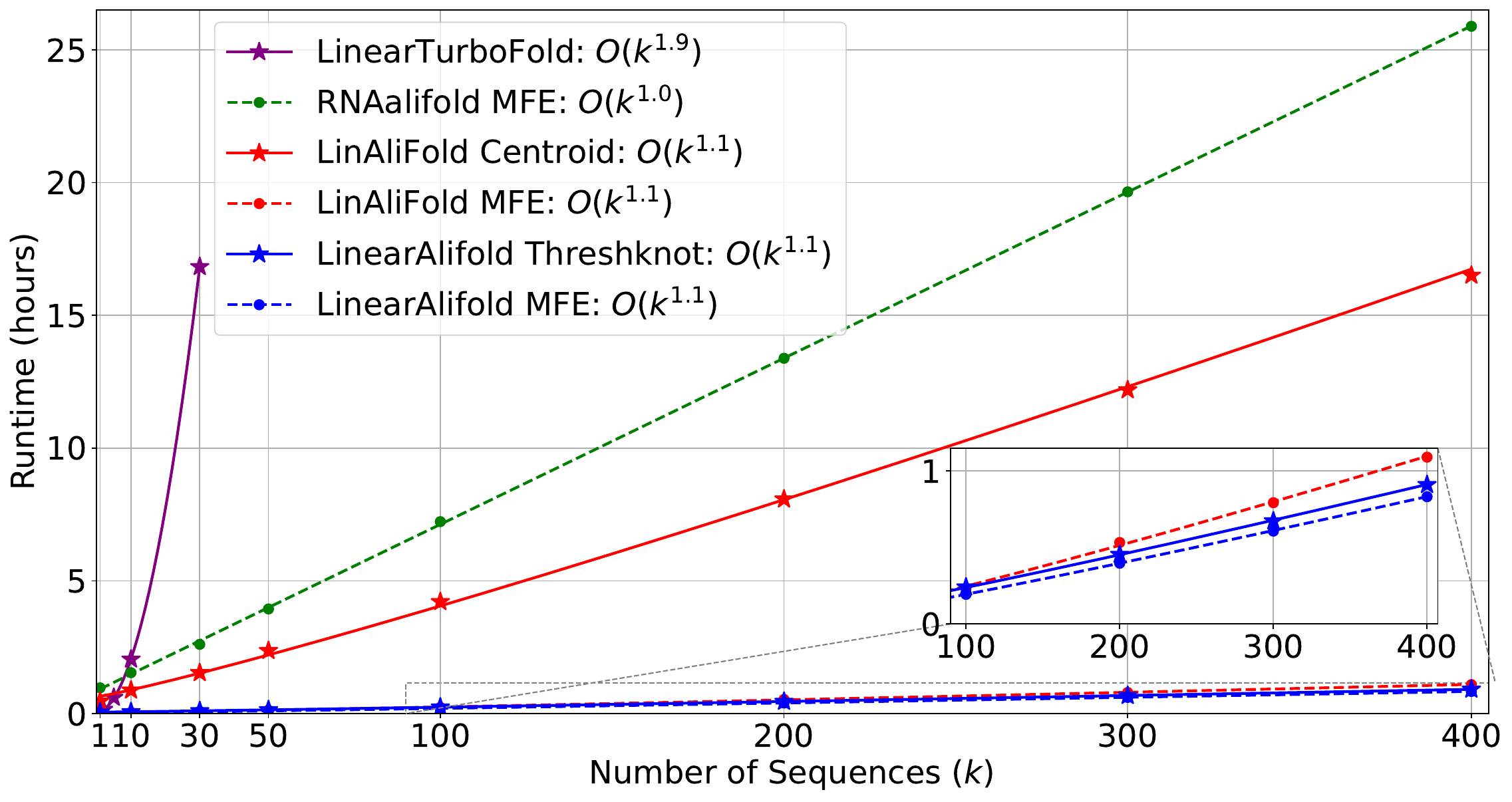} \\ 

\end{tabular}

\vspace{1pt} 

\caption{{\bf A}: Overview of LinearAlifold, which  takes aligned homologous sequences as input to predict consensus MFE structure, consensus partition function, and consensus base-pairing probabilities, which are used in downstream tasks such as Maximum Expected Accuracy (MEA) folding, ThreshKnot folding, and stochastic sampling from the ensemble.
{\bf B}: Runtime of various tools against sequence length ($n$) 
for $k=30$.
{\bf C}: Runtime of various tools against the number of sequences ($k$) for $n\simeq 30,000 \nts$.
}
\label{fig:framework}
\end{figure*}

To alleviate this slow runtime, one can use local folding to predict  structures in linear time, but inevitably
giving up  non-local interactions.
Those base pairs, especially the end-to-end ones, are known to be prevalent in most RNAs \cite{clote+:2012,lai+:2018}.
In particular, the base pairs between the 5' and 3' untranslated regions (UTRs) of  \sarscovtwo,
across $\sim 30,000$ nucleotides,
are found by both purely experimental methods \cite{ziv+:2020} and  purely computational ones \cite{li+:2021}.
How can we achieve linear runtime without without sacrificing long-distance pairs?


Here we report \linearalifold, an efficient tool for consensus structure prediction that scales linearly
with {\em both} the sequence length ($n$) and the number of aligned sequences ($k$)
without any constraints on pair distance,
building upon on our previous work \linearfold~\cite{huang+:2019} and \linearpartition~\cite{zhang+:2020} 
for single sequence folding (Fig.~\ref{fig:framework}A).
Being orders of magnitude faster than \rnaalifold, 
our work can fold hundreds of full-length coronavirus genomes under an hour 
and can recover end-to-end pairs. 
For example, it takes only 0.7 hours to fold the above-mentioned $k=400$ \sarscov sequences,
compared to 25.7 hours by \rnaalifold ($\sim36\times$ speedup).
Meanwhile, \linearalifold significantly outperforms \rnaalifold in structure prediction accuracy compared to 
a database of known structures of homologous sequences~\cite{tan+:2017} (Fig.~\ref{fig:accuracy}). 
More importantly, 
\linearalifold's predictions on hundreds of \sarscov genomes 
(under an hour) 
correlate better with the experimentally guided structures~\cite{huston+:2021,ziv+:2020}
than \rnaalifold's (over a day)
(Fig.~\ref{fig:covid}).
In addition to MFE folding, \linearalifold also supports partition function, base-pairing probabilities, ensemble-based structure prediction methods MEA and Threshknot, and stochastic sampling,
all of which take under an hour on hundreds of \sarscov variants
(which \rnaalifold fails due to overflow).

\linalifold~\cite{fukunaga+hamada:2022} is another 
linear-scaling tool for consensus folding, developed roughly in parallel with our initial version but published earlier.\footnote{Our initial \href{https://arxiv.org/abs/2206.14794}{arXiv preprint (2022)} was discussed in their work.}
Like our work, \linalifold is also built upon our previous work 
\linearfold and \linearpartition, and
thus also achieves linear runtime with both $n$ and $k$.
Unlike our work, their partition-function mode uses
CentroidFold~\cite{hamada+:2009}
and is much slower than our tool, especially with large $k$
(for example, for $k=400$ \sarsr genomes, ours takes 0.8 hours compared to their 16.4 hours, or $\sim20\times$ speedup). 
In fact,  our partition function mode is even faster than their MFE mode (Fig.~\ref{fig:framework}C) thanks to the use of LazyOutside \cite{huang+:2024} (see Methods).
In addition, our tool supports MEA, ThreshKnot (thus our output can contain pseudoknots), and stochastic sampling, none of which is available in their tool.
Moreover, we support two different Turner-style energy models,
the Vienna model  in \rnaalifold and the BL* model~\cite{andronescu+:2010}, while \linalifold only supports the latter. 
More importantly, we also built an easy-to-use web server at \url{http://linearfold.org/linear-alifold}.
 


\vspace{-0.3cm}
\section*{Results}\label{sec:results}

Like \rnaalifold, our \linearalifold also takes a multiple-sequence alignment (MSA) as input (Fig.~\ref{fig:framework}A) and outputs an MFE consensus structure or a consensus partition function.
The scoring function in these systems is a combination of
thermodynamic free energies and sequence covariation scores~\cite{hofacker+:2002,bernhart+:2008} (see Methods).
We employed the beam pruning heuristic to reduce the complexity from cubic runtime (against $n$) to linear time, inspired by \linearfold~\cite{huang+:2019}.
The basic idea of the heuristic algorithm is, at each step $j$, we only keep the $b$ top-scoring states and prune the other ones, which are less likely to be part of the optimal final structure.
This approximate search algorithm helps reduce the time complexity from $O(kn^3)$ to $O(knb^2)$.
In the MFE mode, we further reduced the time complexity to $O(knb \log b)$ following the $k$-best parsing idea \cite{huang+chiang:2005}.
The default beam size is 100, following \linearfold and \linearpartition.
Thus, we reduced the time complexity from $O(kn^3)$ (\rnaalifold) to $O(kn)$ (\linearalifold). 

In the partition function mode, \linearalifold computes
in $O(knb^2)$ time
 the consensus partition function in an ``inside phase'',
which is followed by an ``outside phase'' to compute the consensus base-pairing probabilities (BPPs) (Fig.~\ref{fig:framework}A).
Normally, the outside phase takes the same amount of
time as  inside,
but we employ our (unpublished)  technique  LazyOutside~\cite{huang+:2024}
which takes only $\sim1.5\%$ of the inside time, making inside-outside calculation almost as fast as inside only (and similar to the MFE mode); see Methods for details.
Our tool supports two BPP-based structure prediction methods, MEA and ThreshKnot,  both of which are more accurate than  MFE.
From the consensus partition function, our tool also supports alifold-aware stochastic sampling, based on our LazySampling algorithm.
These sampled structures are useful to ``visualize'' the Boltzmann ensemble, and can be used to compute the accessibility of arbitrary regions~\cite{li+:2021}.

\linearalifold supports two energy models,
Vienna  (as in \rnaalifold) and  BL* (as in \linalifold). 
The latter is our default  model
which generally has higher accuracy on our COVID benchmark
(note that COVID data is disjoint from BL*'s training set).

\subsection*{Scalability}


To demonstrate the scalability of our work, we prepared a set of RNA sequences that contains 8 families ($n\simeq1,600 \nts$ or less) from RNAStralign~\cite{tan+:2017}, 23s rRNA ($n\simeq3,300 \nts$) from the Comparative RNA Web (CRW) site~\cite{Cannone+:2002}, and long sequences (from $\sim9,800 \nts$ to $\sim30,000 \nts$) from three viruses\footnote{HIV 
($n\simeq9,800 \nts$), RSV (Respiratory syncytial virus, $n\simeq15,000 \nts$), and \sarscov ($n\simeq30,000 \nts$) genomes} from NCBI and GISAID\footnote{\url{www.ncbi.nlm.nih.gov} and \url{www.gisaid.org}}.
We used MAFFT~\cite{katoh+:2013} (with \verb|--auto|) to align the input sequences.

Figs.~\ref{fig:framework}B--C compare the  runtime of
three align-then-fold tools (\rnaalifold, \linalifold, and \linearalifold) and one iterative align-and-fold tool (\linearturbofold).
As shown in Fig.~\ref{fig:framework}B, for a given $k$ (here $k=30$), \linearalifold scales linearly with sequence length $n$ and is substantially faster than \rnaalifold
(which scales roughly cubically with $n$)
 under either MFE or  partition function modes. 
In the MFE mode, on the SARS-CoV family, 
\linearalifold 
is $\sim40.8\times$ faster than \rnaalifold
(3.8 min.~vs.~2.6 hours)
In the partition function mode, \rnaalifold cannot scale to $n>14,000 \nts$ for $k=30$ 
due to overflow.
On the HIV family ($\simeq9,800 \nts$),
\linearalifold's ThreshKnot mode is $\sim120\times$ faster 
than \rnaalifold's MEA mode (1.2 min.~vs.~2.5 hours) 

We also tested  runtime against the number of homologs ($k$) using 
SARS-CoV ($n\simeq30,000 \nts$) (Fig.~\ref{fig:framework}C).
Here all three align-then-fold tools (\rnaalifold, \linalifold, and \linearalifold) scale linearly with $k$, but the iterative align-and-fold tool \linearturbofold scales quadratically with $k$,
making it only feasible on $\sim30$ \sarsr genomes.
For $k=400$, \rnaalifold requires more than a day (25.7 hours)
while our tool only needs 0.7 hours ($\sim 36.3\times$ speedup).
In addition, \rnaalifold partition function mode fails to run 
on \sarscov due to overflow.


Although \linalifold also scales linearly with both $n$ and $k$,
our tool is still substantially faster, especially in the partition function mode. 
This is due to two reasons:
(a) we employ LazyOutside~\cite{huang+:2024} which reduces the outside phase to just 1--2\% of the inside phase, bringing a $\sim2\times$ speedup; and
(b) their partition function mode uses CentroidFold, which mixes consensus BPPs with individual single-sequence BPPs (i.e., calling \linearpartition $k$ times).
As a result, on $k=400$ \sarsr genomes, our \linearalifold ThreshKnot takes 0.8 hours compared to their 16.4 hours ($\sim20\times$ speedup). 
Actually, our partition function mode is even faster than their MFE mode (0.8 vs.~1 hour(s)).

\begin{figure}[!t]
\hspace{-0.4cm}
\begin{tabular}{c}
{\panel{A}}\\
\includegraphics[width=\linewidth]{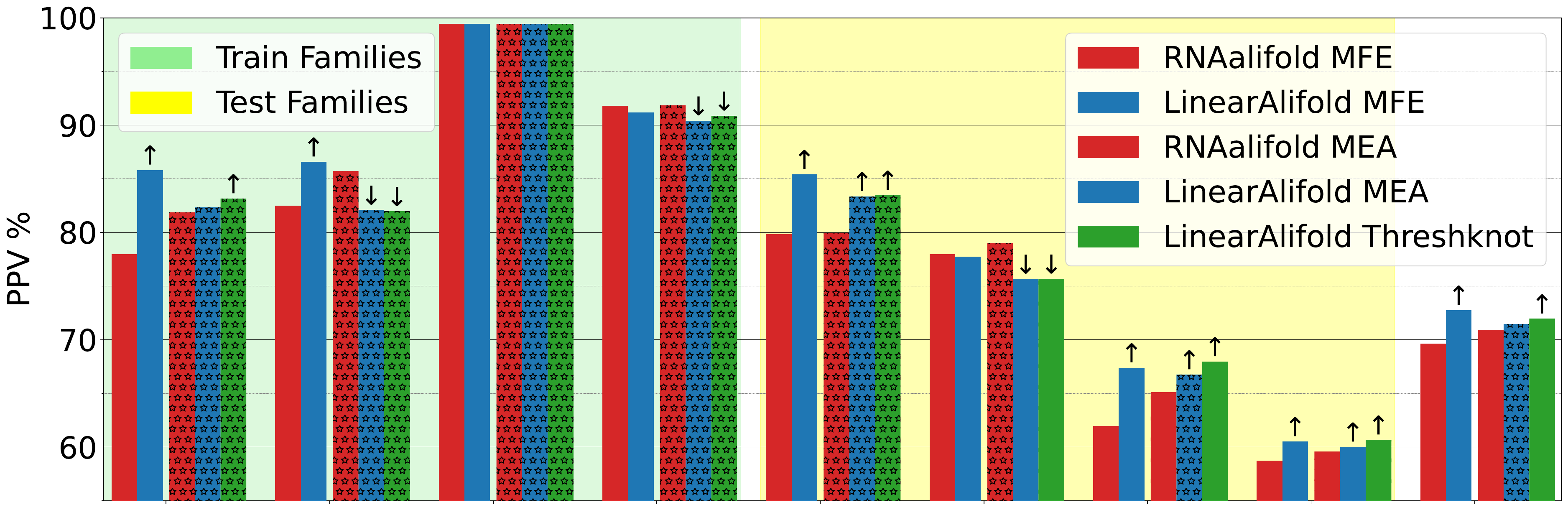}\\
{\panel{B}}\\
\includegraphics[width=\linewidth]{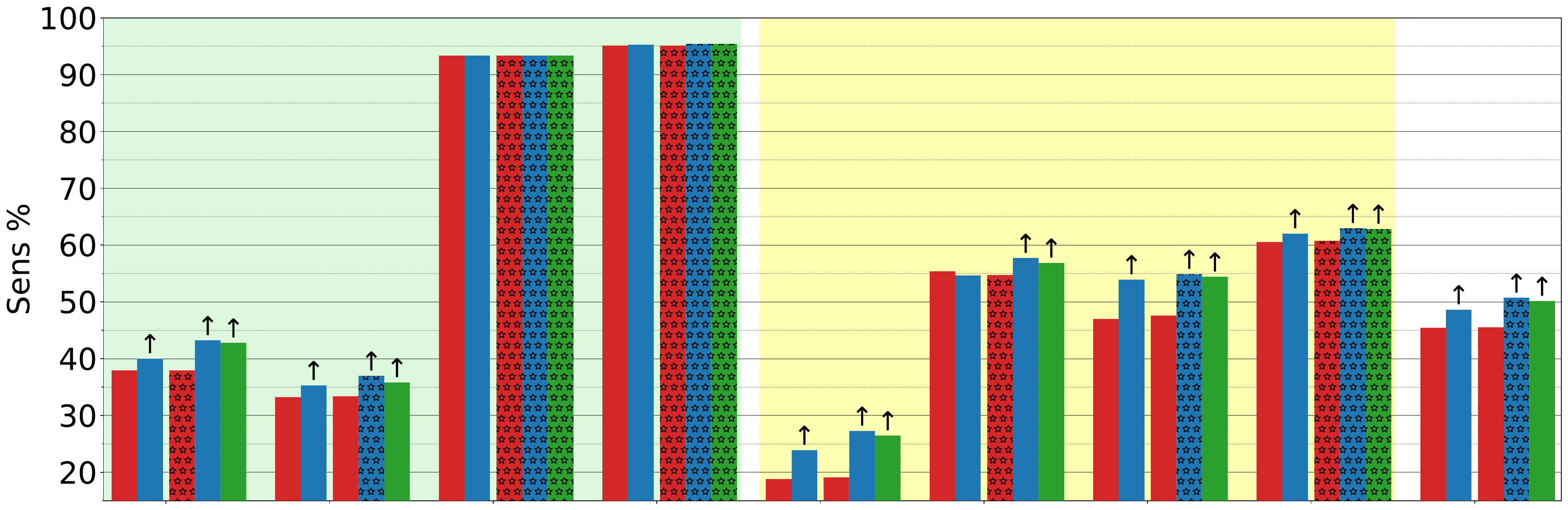}\\
{\panel{C}}\\
\includegraphics[width=\linewidth]{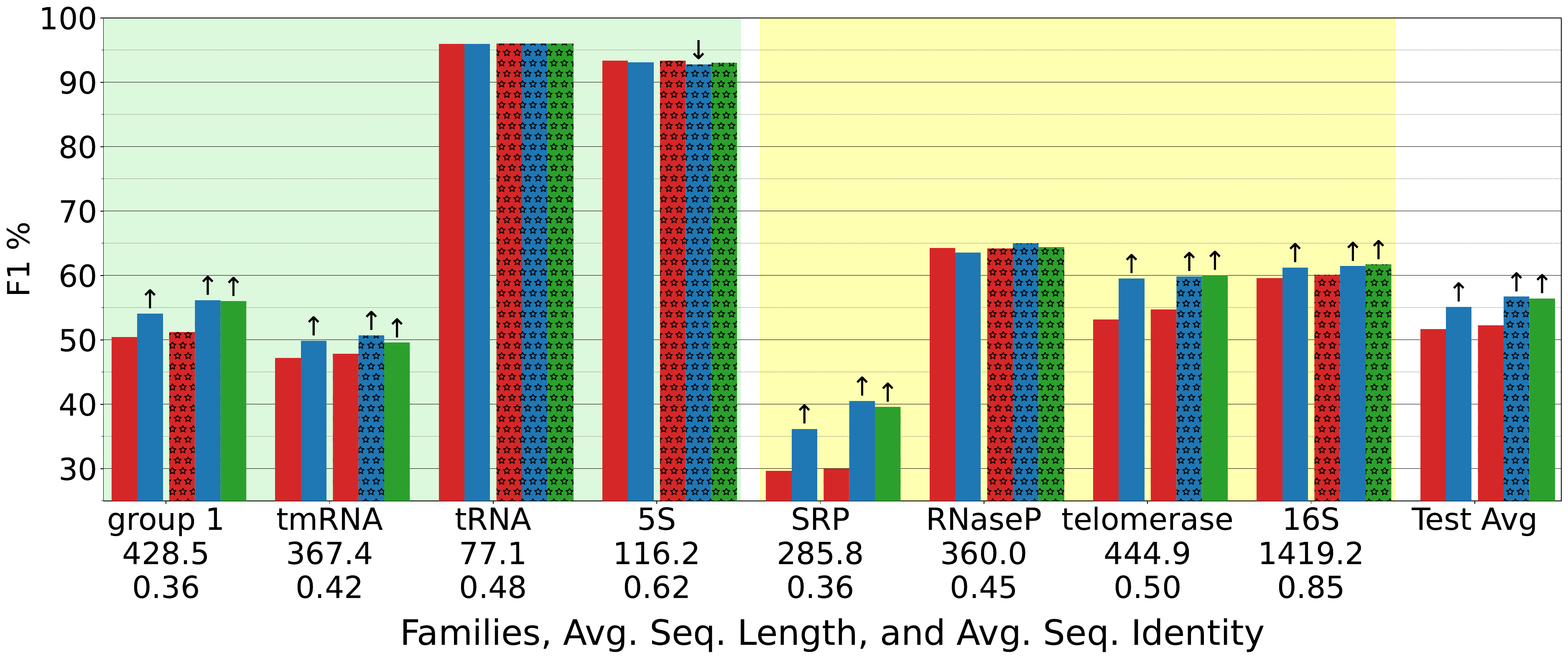} 
\end{tabular}
\caption{Accuracy comparisons between \rnaalifold and \linearalifold; each family has 10 samples and each sample is has $k=30$ homologs. 
Statistical significance (two-sided) is marked as `$\uparrow$' if \linearalifold is significantly better, 
or `$\downarrow$` if \rnaalifold is significantly better ($p \!<\!0.05$). See also Fig.~\ref{fig:si-accuracy-10}.
\label{fig:accuracy}
}
\vspace{-0.2cm}
\end{figure}
\subsection*{Accuracy}

We compared the accuracies of secondary structure prediction using 
the RNAStralign database~\cite{tan+:2017}, which have well-determined secondary structures of RNA homologs
for eight  families (Fig.~\ref{fig:accuracy}). 
For each family, we take 10 samples, each of which contains $k=30$ sequences. These sequences in each sample were first aligned using MAFFT (\verb|--auto|) before being fed into \rnaalifold and \linearalifold (both using the Vienna energy model).
Following \linearturbofold, we used the first four families 
(tRNA, 5S rRNA, tmRNA, and Group I Intron)
to tune the hyperparameters (see Methods),
so the ``Test Avg'' columns include the remaining four families
(SRP, RNaseP, telomerase, and 16S rRNA).


\begin{figure*}[!t]
\hspace{-0.4cm}
\begin{tabular}{ccc}

& \multicolumn{1}{c}{\large{Vienna Energy Model (A, C, E)}} & \multicolumn{1}{c}{\large{BL* Energy Model (B, D, F)}} \\[5pt]

& \multicolumn{1}{c}{\textbf{A}} & \multicolumn{1}{c}{\textbf{B}} \\
& \includegraphics[width=.48\linewidth]{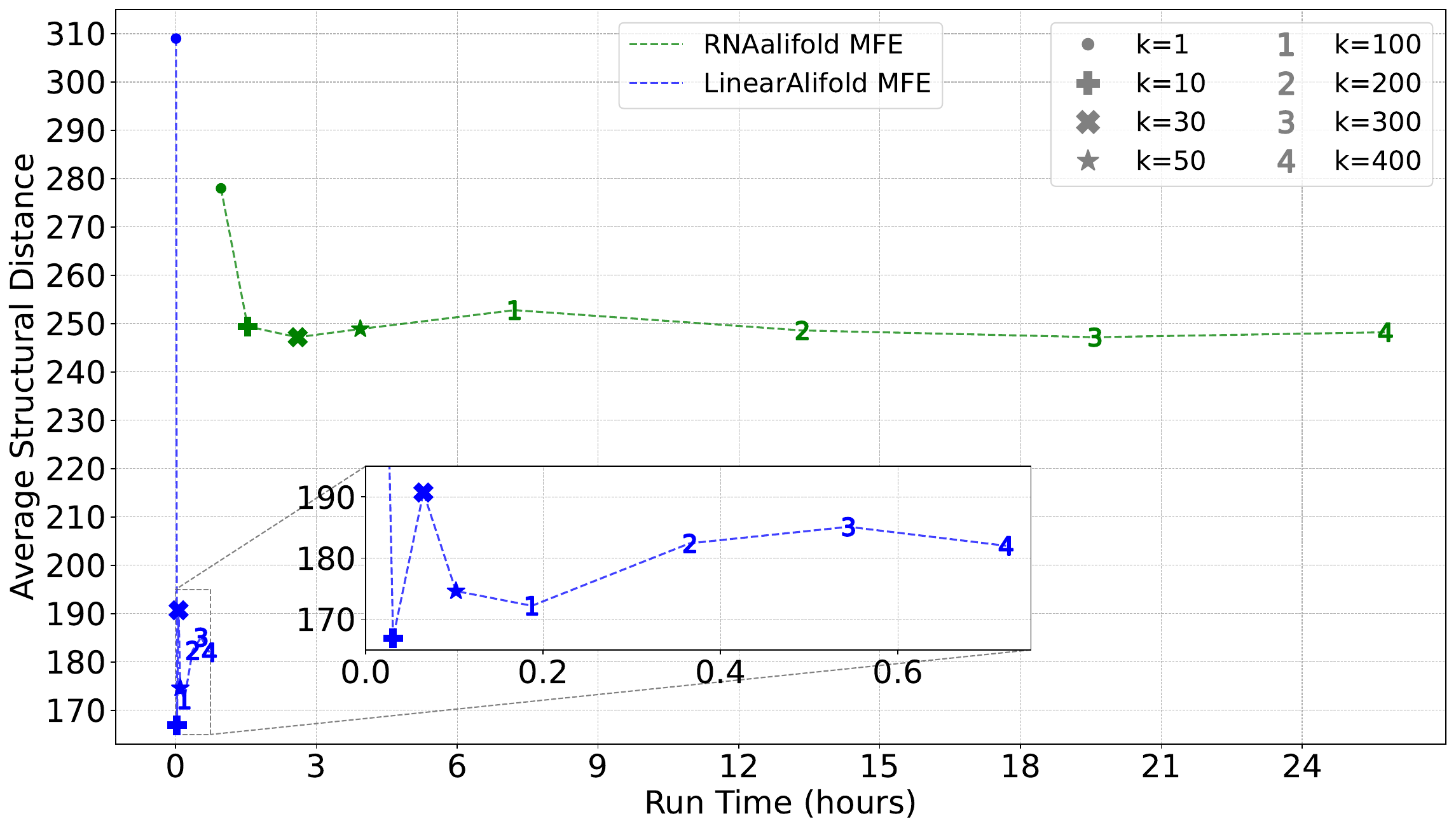}
& \includegraphics[width=.48\linewidth]{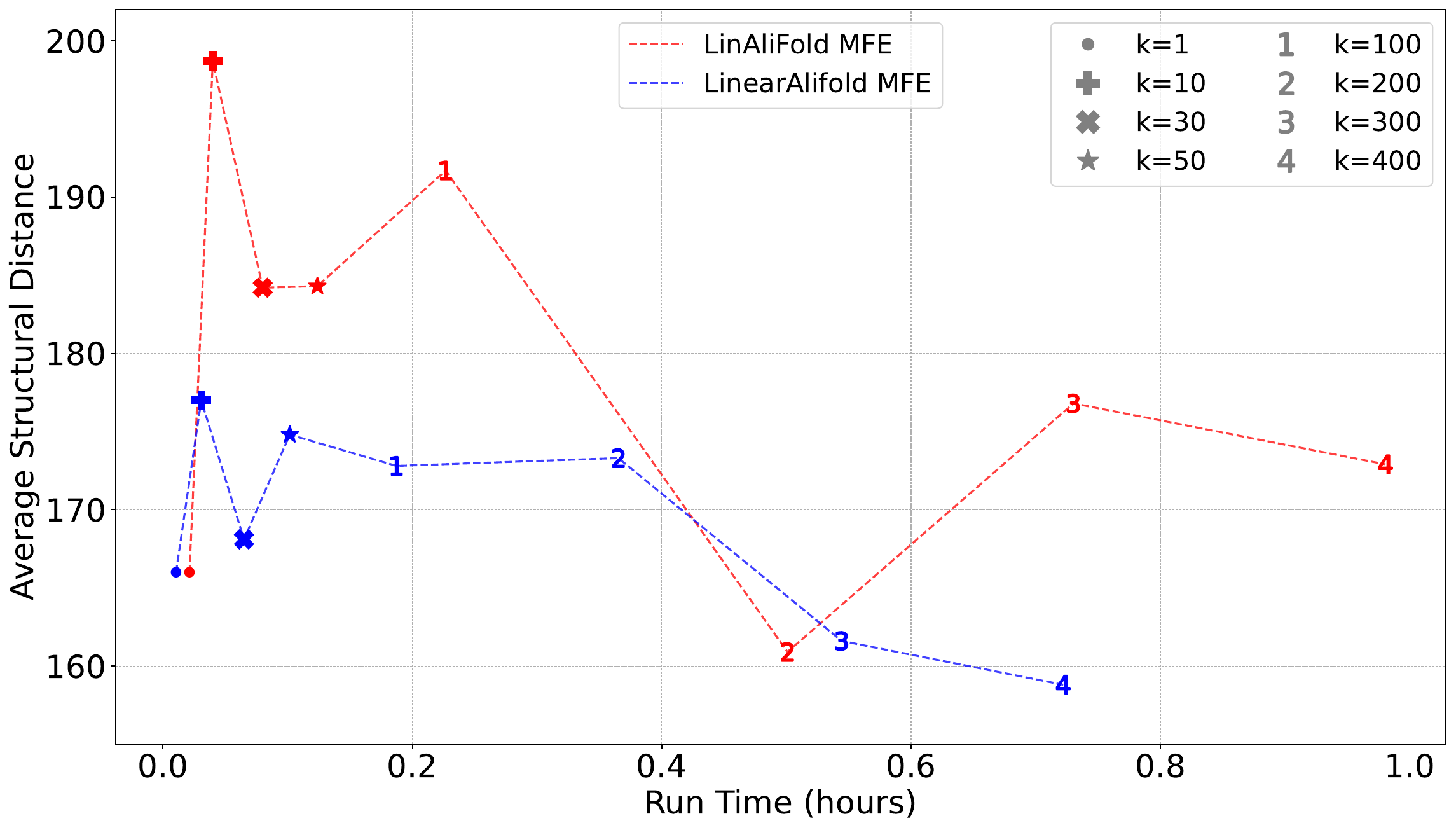} \\

& \multicolumn{1}{c}{\textbf{C}} & \multicolumn{1}{c}{\textbf{D}} \\
& \includegraphics[width=.48\linewidth]{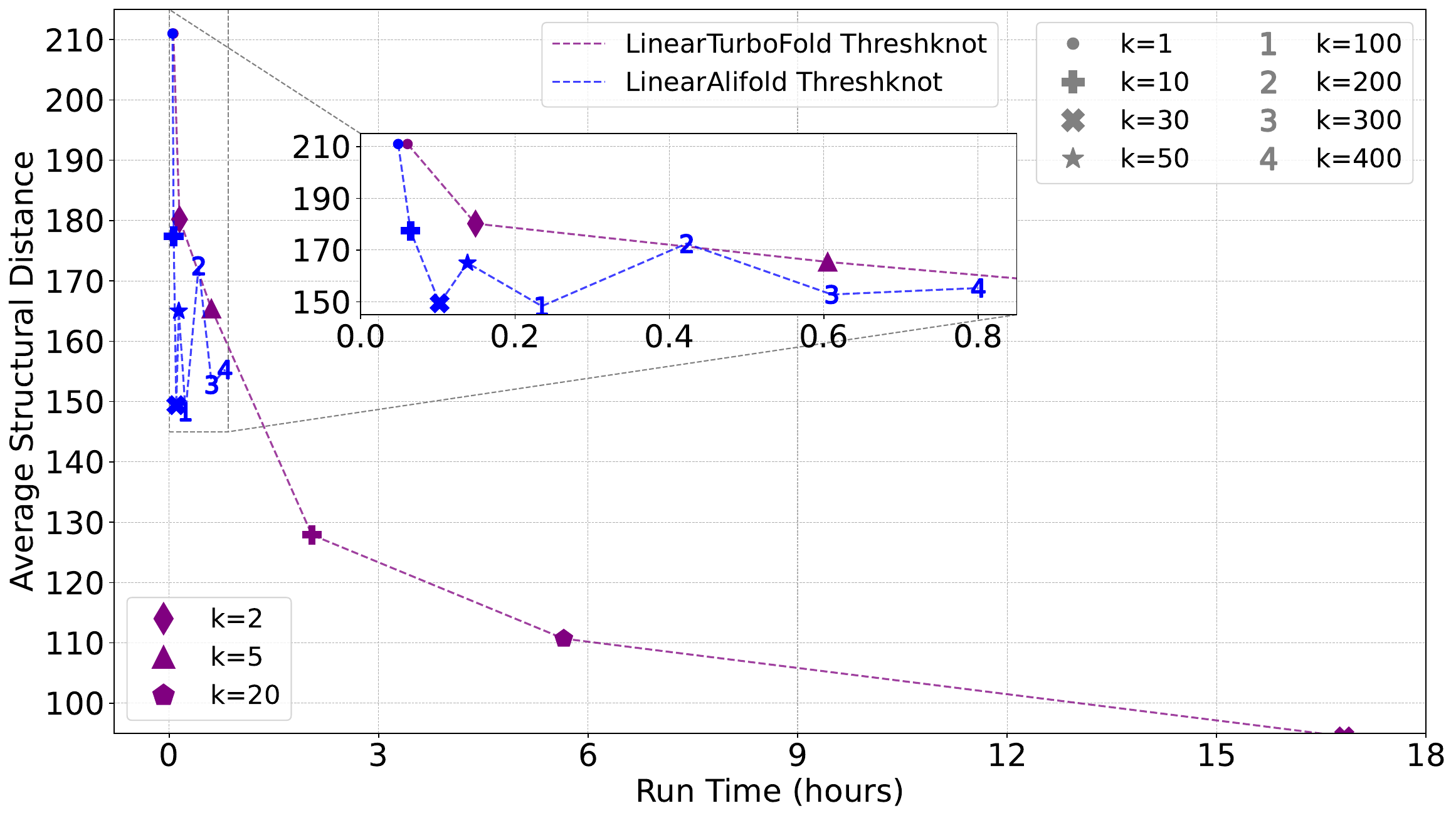} 
& \includegraphics[width=.48\linewidth]{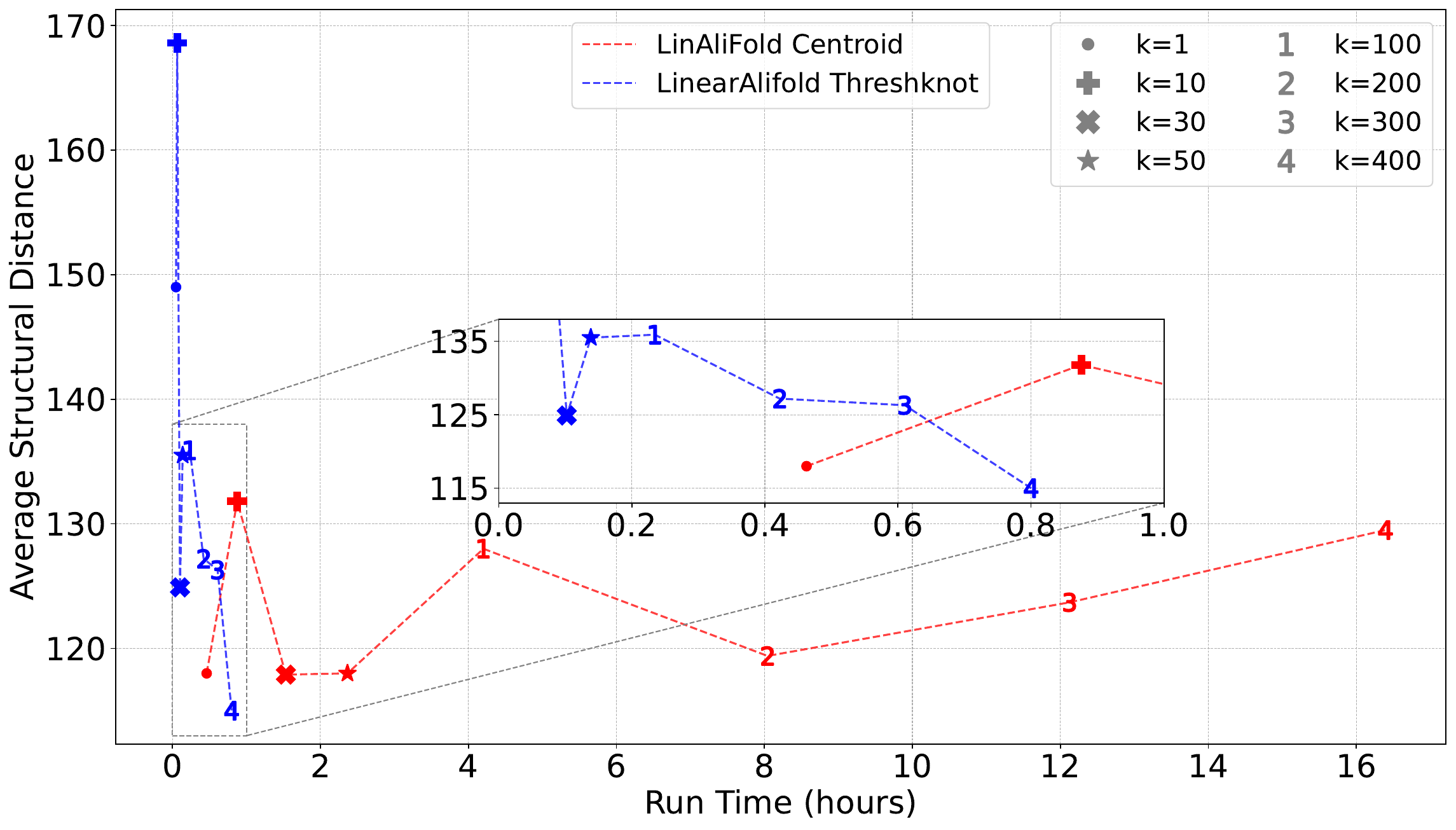} \\

& \multicolumn{1}{c}{\textbf{E}} & \multicolumn{1}{c}{\textbf{F}} \\
& \includegraphics[width=.48\linewidth]{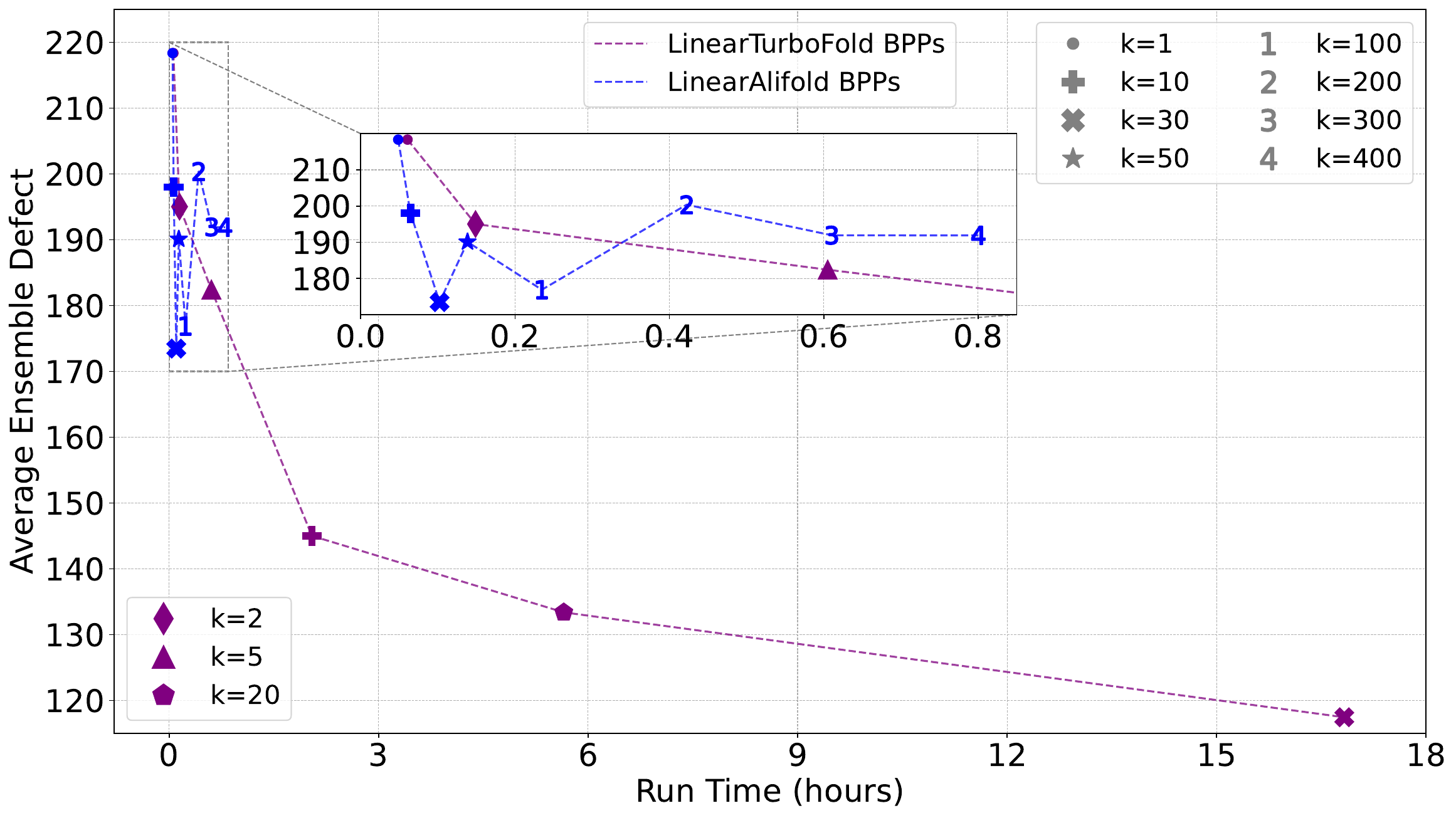} 
& \includegraphics[width=.48\linewidth]{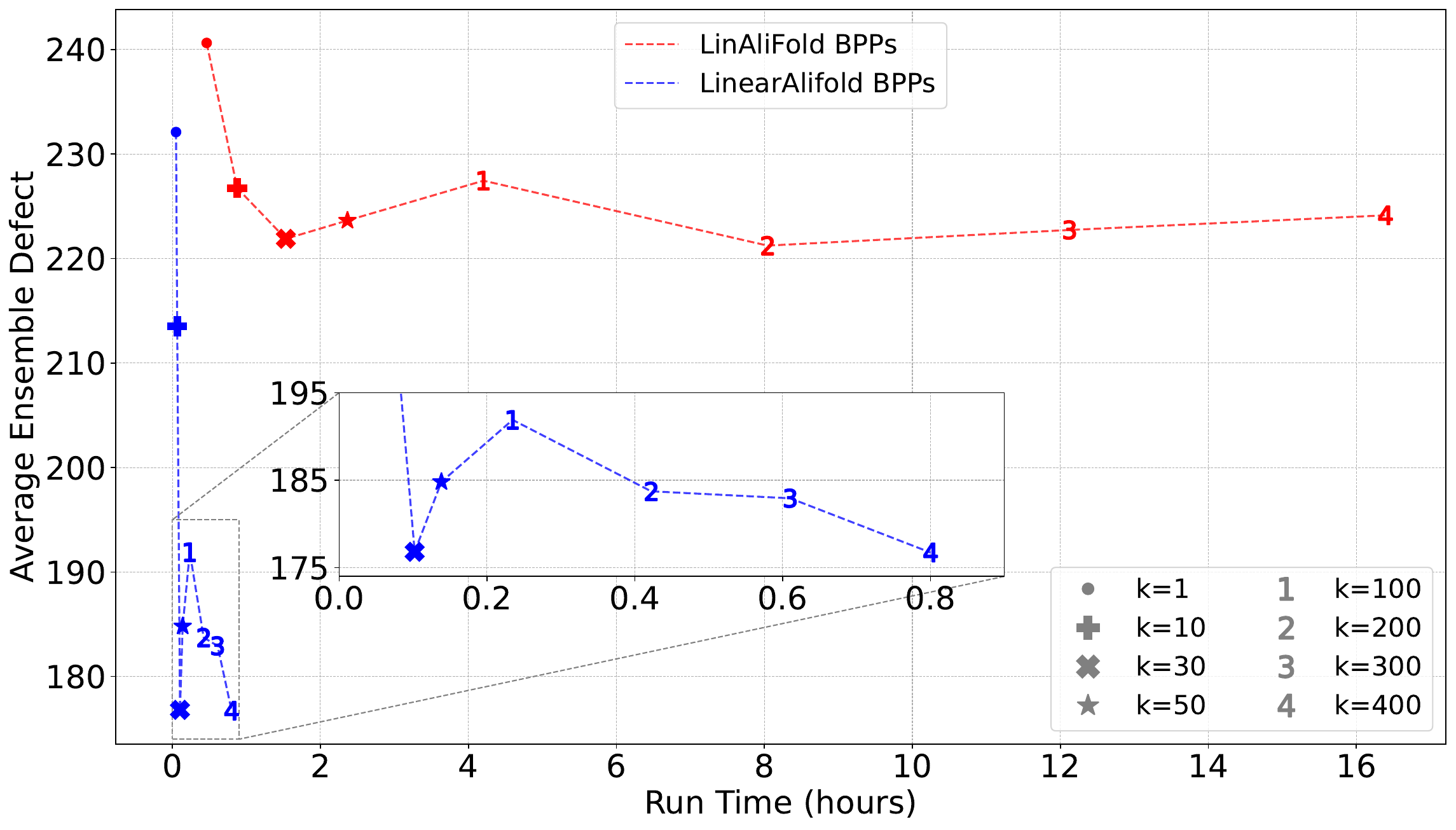} \\

\end{tabular}


\caption{Structural distance and ensemble defect against run time for different energy models and different methods. The curves show the mean values over 10 samples for each $k$
A--B: MFE prediction. C--D: partition-based structure prediction.
E--F: ensemble quality.
See Fig.~\ref{fig:covid-si} for another version which shows more statistics of each 10 samples
and uses $k$ as the x-axis.}
%
%
\label{fig:covid}
\end{figure*}

In terms of F1 score,
\linearalifold's MFE and MEA modes significantly outperform the corresponding modes of \rnaalifold 
on all test families,
and \linearalifold's ThreshKnot mode
significantly outperforms \rnaalifold's MEA mode on almost all test families except for RNaseP (two-sided significance test~\cite{Aghaeepour+Hoos:2013}).  
This high accuracy of LinearAlifold over RNAalifold is 
expected, and is due to the beam search in the former, which is inherited from LinearFold~\cite{huang+:2019}. As we showed in our LinearFold paper, although beam search introduces minor search errors and returns suboptimal structures in terms of the scoring function, it nevertheless makes the search more robust locally (since the scoring function is never perfect), which translates to slightly better accuracy compared to ground-truth structures. We observe this phenomenon over and over in our previous work LinearFold, LinearPartition~\cite{zhang+:2020}, LinearSampling~\cite{zhang+:2022}, LinearCoFold~\cite{zhang+:2023}, and LinearTurboFold~\cite{li+:2021}, as well as our earlier work in natural language parsing \cite{huang+sagae:2010} which gave rise to LinearFold, so this is a universal phenomenon.

Since the BL* energy model is trained on structures which overlap with our benchmark, 
it overfits on it. Thus we do not include our results with BL* (nor  a comparison with \linalifold using the same energy model). Figs.~\ref{fig:si-accuracy-10} and~\ref{fig:si-accuracy-20} compare more systems including LinearTurboFold, LinAlifold, and single-sequence folding. 

Note that align-then-fold systems (RNAalifold, LinearAlifold, and LinAliFold) tend to be inaccurate for low sequence indentity families (e.g., SRP and group 1) and tend to be more accurate for high sequence identity families (e.g., 16S rRNA).



\begin{figure*}[!t]
    \centering
    \begin{tabular}{ccc}
        \small{{\bf A:} LinearAlifold Vienna BPP}
        &
        \small{{\bf B:} LinearAlifold BL* BPP}
        &
        \small{{\bf C:} LinearTurboFold Vienna BPP}
        \\[0.2cm]
        \includegraphics[width=0.3\textwidth]{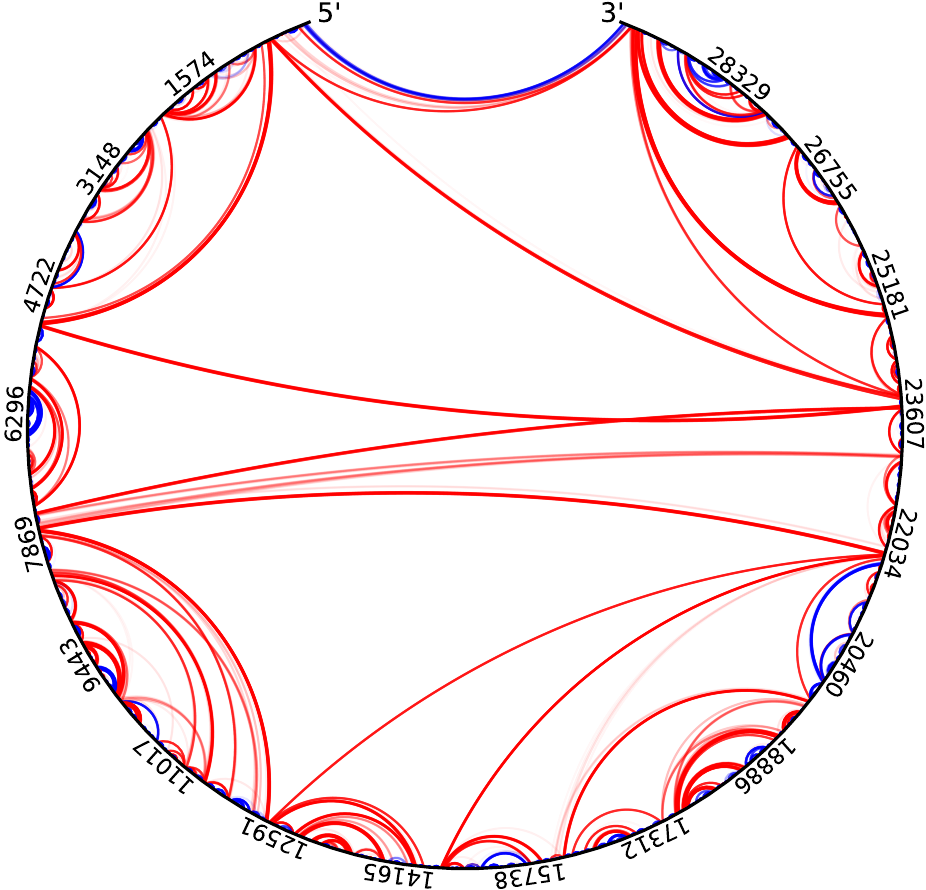}
        &
        \includegraphics[width=0.3\textwidth]{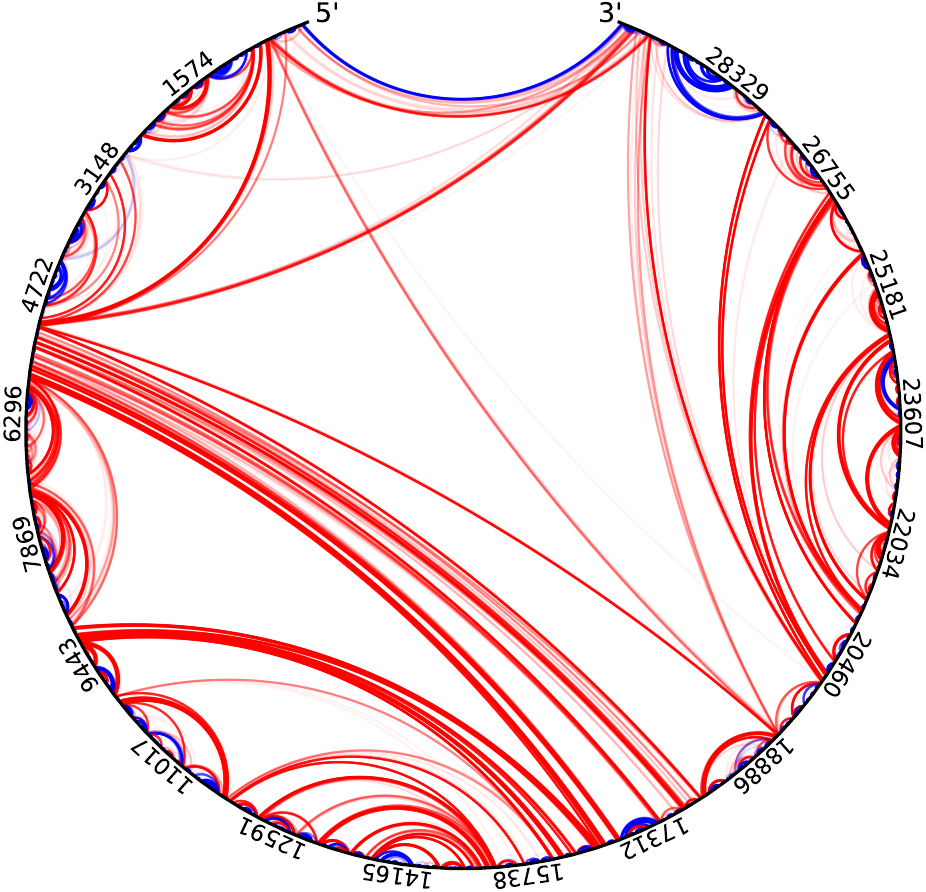}
        &
        \includegraphics[width=0.3\textwidth]{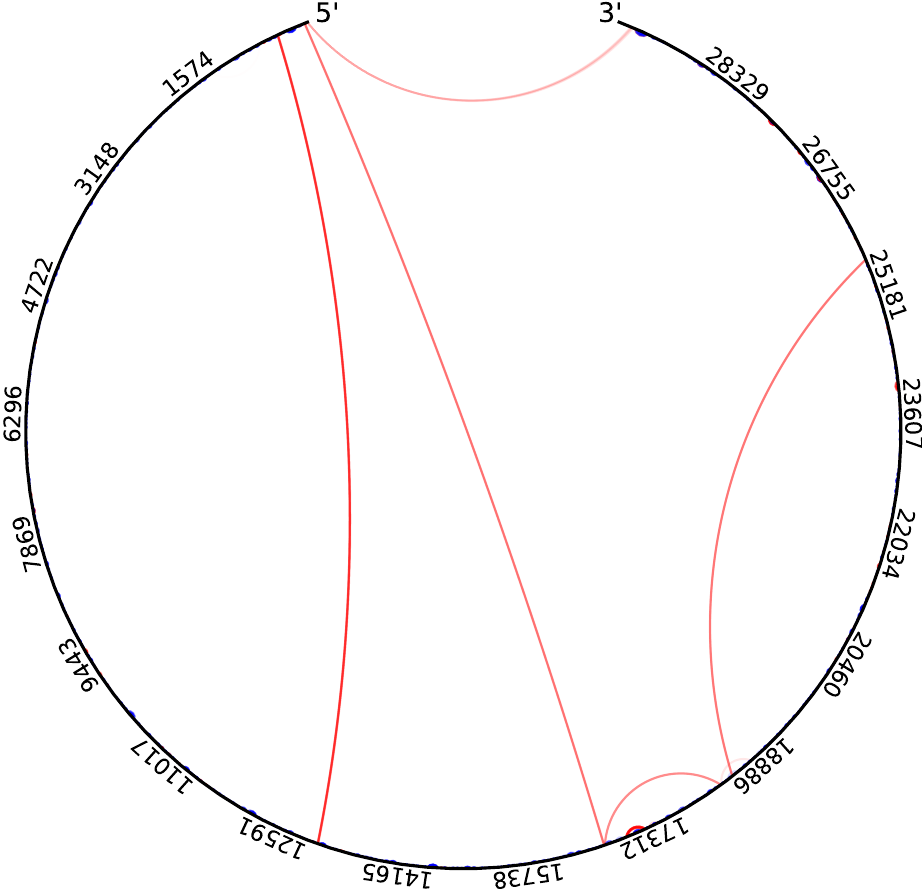}
        \\[0.1cm]
        \small{{\bf D:} Stochastic Sampling}
        &
        \small{{\bf E:} Stochastic Sampling}
        &
        \small{{\bf F:} Stochastic Sampling}
        \\
        \includegraphics[width=0.3\textwidth]{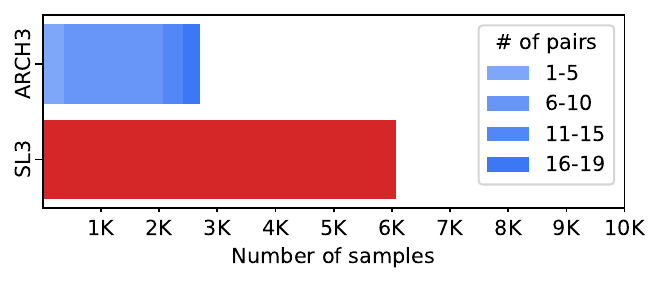}
        &
        \includegraphics[width=0.3\textwidth]{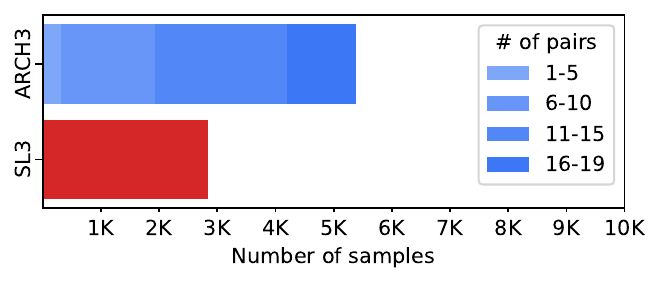}
        &
        \includegraphics[width=0.3\textwidth]{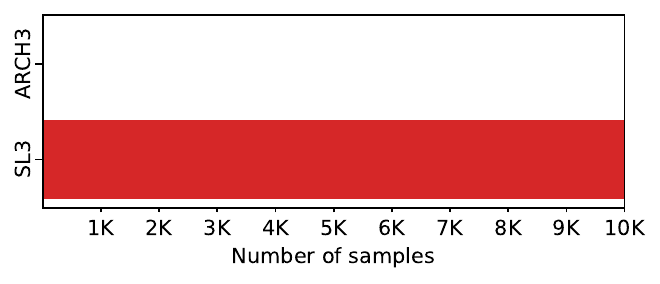}
        \\
    \end{tabular}\\[0.1cm]
    \hspace{-0.3\textwidth}{\bf G} \hspace{0.5\textwidth} {\bf H}\\
    {\includegraphics[width=1\textwidth]{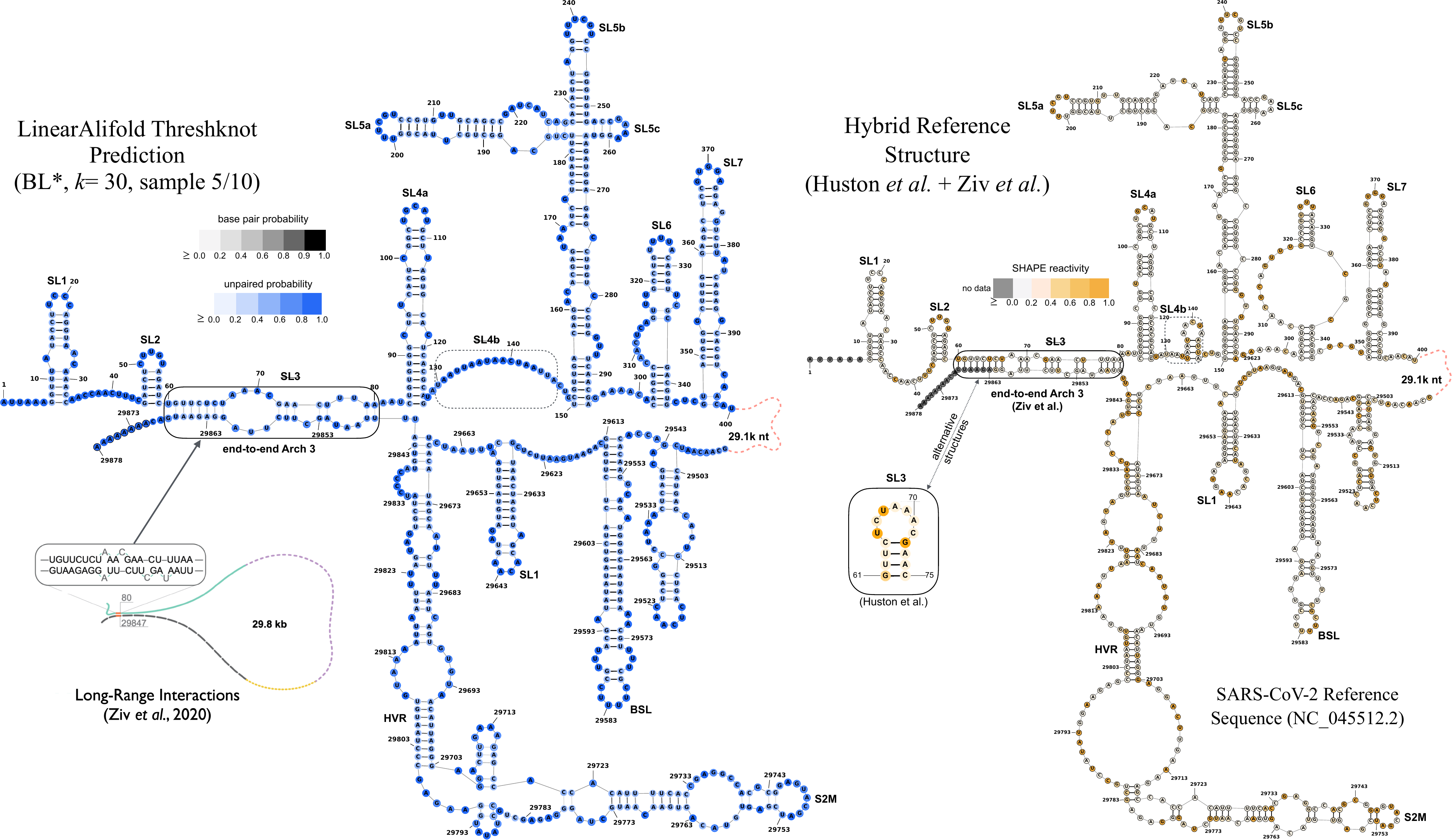}}
    \caption{Visualizations of structure predictions on $k=30$ SARS-CoV genomes (A--G) compared with the experimentally-guided hybrid structure (H).
    A--C: Circular plots of base-pairing probabilities (BPPs) from LinearAlifold (two energy models) and LinearTurboFold on $k=30$ genomes (sample 5/10). Blue arcs are consistent with at least one range from Ziv et al.~\cite{ziv+:2020}, while red arcs are not supported by any such range. The darkness of the arcs indicates pairing probability. D--F: stochastic sampling statistics (over 10,000 structures) between the competing global (arch 3 from Ziv et al.) and local (SL3 from Huston et al.~\cite{huston+:2021}) structures. G: the 5' and 3' UTR structures of LinearAlifold (BL*) ThreshKnot prediction, with shades of blue for unpaired probabilities of each nucleotide and shades of black for pairing probabilities for each pair. H: the reference hybrid structure based on Huston et al.'s SHAPE-guided model but with the end-to-end arch 3 from Ziv et al.~replacing SL3.}
    \label{fig:hybrid}
\end{figure*}

\subsection*{Consensus Structure Prediction in \sarscovtwo and \sarsr Betacoronaviruses}

It is known that conserved structures across mutations are critical for viruses to maintain their functions to survive.
Thus, these conserved regions could be potential targets for diagnostics and therapeutics~\cite{nawrocki+:2013,brown+:1992,ritz+:2013}. 
To model consensus structures for  \sarscovtwo and \sarsr betacoronaviruses, 
for each $k$ ranging from 10 to 400,
we sampled 10 sets of diverse sequences (see Methods for details), 
and used MAFFT \verb|--auto| to generate 10 MSAs for each $k$. 
Following \linearturbofold \footnote{The \linearturbofold paper \cite{li+:2021} built a dataset of 25 SARS-CoV genomes: 16 \sarscovtwo plus 9 \sarsr sequences.},
the ratio of the number of \sarscovtwo to the number of \sarsr genomes remains 6 to 4 in all samples. 

To evaluate the reliability of \linearalifold's prediction on \sarscovtwo, 
we compared the predicted  structure with experimental studies~\cite{huston+:2021,ziv+:2020} for the well-known 5' and 3' UTR regions. 
Huston et al.~\cite{huston+:2021} modeled secondary structures guided by the chemical probing data, 
but used a local folding method for prediction because the sequence length of \sarscovtwo is out of reach of most algorithms. 
As a result, long-range interactions were fully abandoned in their prediction,  
which are critical for regulating the viral transcription and replication pathways~\cite{lai+:2018,ziv+:2020}.
To overcome this issue, 
we further involved a purely experimental study of Ziv et al.~\cite{ziv+:2020}, 
which can detect long-range interactions between 5' and 3' UTRs. 
Therefore, 
to take into consideration both local and global structures between 5' and 3' UTRs,
we built a hybrid structure model (Fig.~\ref{fig:hybrid}H) by combining 
Huston et al.~and Ziv et al.'s 
work (see  Methods).  

Fig.~\ref{fig:covid} compares the quality of predictions 
from four tools (\rnaalifold, \linalifold, \linearturbofold, and \linearalifold), two energy models Vienna (A/C/E) and BL* (B/D/F), and three modalities (MFE (A--B), partition-based structure prediction (ThreshKnot/Centroid, C--D), and base-pairing probabilities (E--F)). The metrics are structural distance (the number of incorrectly predicted nucleotides) and ensemble defect
(the expected structural distance over the Boltzmann ensemble), both the lower the better
(closer to the above hybrid structure model).
The x-axes in these plots are run time, 
showing the speed advantage of our tool over others.

In Fig.~\ref{fig:covid}A, our MFE
is substantially faster and more accurate than 
\rnaalifold MFE (both with Vienna energy model), and in panel B, our MFE
is noticeably faster and more accurate than \linalifold MFE
(both with BL* energy model).
Next, Figs.~\ref{fig:covid}C--D
compare our tool with \linearturbofold and \linearalifold
in terms of partition-function-based structure prediction
(note that as mentioned before, \rnaalifold's partition function mode does not run on \sarscov genomes).
In Fig.~\ref{fig:covid}C, the iterative align-and-fold tool \linearturbofold achieves substantially better structural distance than our align-then-fold tool, presumably due to folding-aware alignment, but at the cost of much slower run time 
and inability to scale beyond $k=30$.
In Fig.~\ref{fig:covid}D, \linalifold Centroid mode achieves similar structural distance 
as our ThreshKnot mode, but takes $\sim20\times$ more time
due to mixing with single-sequence BPPs.
Finally, Figs.~\ref{fig:covid}E--F are  similar to C--D, but instead of evaluating one predicted structure, they evaluate the quality of the whole ensemble,
measured by the ensemble defect computed using the base-pairing matrix. 
The only difference is that in F, 
\linalifold's ensemble quality (still with the same mixing in D)
is substantially worse than ours,
suggesting that CentroidFold was able to extract a high quality structure from a lower quality ensemble.

Across the board, the BL* column (B/D/F) is consistently better than the Vienna column (A/C/E), so we choose BL* as the default energy model, but the user can change it with a command-line switch.

Fig.~\ref{fig:covid-si} is similar to Fig.~\ref{fig:covid} but uses $k$ as the x-axis, and
 draws the 25-75 quantile boxes (and medians) in addition to the mean curves, since we have 10 samples for each $k$.
Fig.~\ref{fig:huston-only} is similar to Fig.~\ref{fig:covid-si} but uses Huston et al.'s model structure instead of the hybrid structure as the reference.

 \smallskip 
 
To further visualize our predicted structures, we choose one particular sample (\#5/10) for $k=30$ SARS-CoV-2 and SARS-related genomes; this $k$ is chosen because it is the largest for LinearTurboFold to run, and this particular sample is chosen because our LinearAlifold BL* prediction (our default setting) achieves the best structural distance (against the hybrid structure).
Fig.~\ref{fig:hybrid}A--C compare the base-pairing probabilities (BPPs) for three systems: LinearAlifold Vienna model, LinearAlifold BL* model, and LinearTurboFold (Vienna model). Here we use Ziv et al.'s ranges as references, and blue arcs indicate pairings supported by at least one Ziv et al.~arc, and red ones are not supported by any Ziv et al.~arc. We can see that LinearAlifold systems predict many more non-local (long-distance) pairing possibilities, although most of them are incorrect,
and LinearTurboFold mostly predicts local pairings. 
Fig.~\ref{fig:circular-ziv} shows the corresponding ThreshKnot predictions (grouped by pairing distance)
and their precision against Ziv et al.~ranges.
We can see that LinearAlifold's both models predicted about 2,000 non-local pairs ($\ge 100$ \nts), among which 36.4\% of the prediction by BL* model and 32.3\% of the prediction by Vienna model are supported by at least one Ziv et al.~ranges, respectively. LinearAlifold BL* model also predicted 14 end-to-end pairs which are all supported by Ziv et al., whereas the other two systems did not predict any end-to-end pairs.

More interestingly, we would like to further investigate the competition between alternative structures in the Boltzmann ensemble, in particular, the end-to-end arch 3 (from Ziv et al.) vs.~the local SL3 in 5' UTR (from Huston et al.). Fig.~\ref{fig:hybrid}D--F conduct stochastic sampling for LinearAlifold Vienna, LinearAlifold BL*, and LinearTurboFold. Interestingly, LinearAlifold BL* prefers end-to-end arch 3 (but with about 30\% of sampled structures showing SL3), while LinearAlifold Vienna prefers SL3 (about 60\% of sampled structures). LinearTurboFold, however, is 100\% SL3.

Finally, Fig.~\ref{fig:hybrid}G shows the 5' and 3' UTR structure of the LinearAlifold BL* ThreshKnot prediction,
which is very similar to the hybrid reference structure in Fig.~\ref{fig:hybrid}H.
It is also worth noting that, unlike Huston et al.'s experimentally guided model, LinearAlifold BL* ThreshKnot predicts the SL4b region to be single-stranded (Fig.~\ref{fig:hybrid}G),
which is consistent with the experimentally guided structure by Sun et al.~\cite[Fig.~2C]{sun+:2021}. These results, plus the fact that the prediction from the LinearTurboFold paper~\cite[Fig.~3]{li+:2021} has rather weak and different pairs for SL4b, all suggest alternative structures in the ensemble for that region.

\section*{Discussion}\label{sec:discussion}

Considering the fast mutation rate of RNA viruses such as \sarscovtwo, 
accurately identifying conserved regions from homologs is critical to 
develop mutation-insensitive diagnostics and therapeutics. 
Consensus folding algorithms, which can take hundreds of aligned homologs
to predict consensus structure, 
are widely-used for this task.
However, \rnaalifold, the most widely used consensus folding tool,
scales cubically with the sequence length in runtime, 
and is prohibitively slow to analyze long RNAs,
especially \sarscovtwo ($\sim30,000$ $nt$).
To alleviate this issue, we present \linearalifold, 
an efficient tool which scales linearly with both the sequence length ($n$) and the number
of aligned sequences ($k$). 
We confirmed that \linearalifold is orders of magnitude faster than \rnaalifold,
taking less than an hour to fold 400 full-length \sarscov genomes (which takes more than a day for \rnaalifold MFE mode).
We also demonstrated that \linearalifold achieves significantly higher accuracies on a benchmark dataset with known structures.
\linearalifold is also faster than a similar linear-time consensus folding tool \linalifold, 
especially in the partition function mode and for a larger $k$.

\linearalifold has four output modalities: (1) predicting consensus minimum free energy structure (MFE mode); 
(2) predicting the MEA structure based on the consensus BPP;
(3) predicting the ThreshKnot structure  based on the consensus BPP; and
(4) stochastically sampling structures from the consensus partition function.
All these modes can be applied to hundreds of aligned \sarscovtwo homologs, 
while \rnaalifold can only handle the MFE mode for such MSAs due to overflow,
and \linalifold only supports (1) and a variant of (3) (CentroidFold).
\linearalifold's prediction on \sarscovtwo correlates better with experimentally-guided structures
than \rnaalifold's or \linalifold's, yet takes substantially less time.

\linearalifold is a general algorithm and can also be applied to analyze other long RNA viruses, such as HIV, WNV (West Nile Virus), and Ebola.
Finally, we built a web server which will be useful for biologists.

\section*{Methods}\label{sec:methods}
 \captionsetup[figure]{font={stretch=1}} 
 \renewcommand{\thefigure}{S\arabic{figure}}
 \renewcommand{\thetable}{S\arabic{table}}

\subsection*{Scoring function of \rnaalifold and \linearalifold}
Following \rnaalifold, for a set of $k$ aligned sequences $S$, our scoring function takes into consideration  both a thermodynamic energy model and a sequence covariation score
$\gamma(i,j,S)$ to evaluate the corresponding alignment column pair $(i,j)$'s compensatory mutations:
\[
\score(S,\, y)  = \frac{1}{k} \Big[\sum_{s\in S} \Delta G(s, y) \, + \, \beta\!\!\sum_{(i,j)\in y} \gamma(i, j, S)\Big]
\]
where  $y$ is a consensus secondary structure, $\Delta G(s, y)$ is the free energy of sequence $s$ folded into structure $y$ (when mapping consensus structure $y$ on to an individual sequence $s$, we remove the pairs in $y$ that are not pairable on $s$), and $\gamma(i,j,S)$ is the {\em (base pair) conservation score} that evaluates the corresponding alignment columns with respect to evidence for base pairing
\[\hspace{-0.5cm}
\gamma(i,j, S) =\frac{1}{k}\gamma'(i,j, S)+\delta \sum_{s \in S} 
\begin{cases}
0  & \textrm{ if } (s_i,s_j) \in \pairables\\
0.25 &  \textrm{ if } (s_i, s_j) = (-, -)\\ 
1 & \textrm{ otherwise}
\end{cases} 
\label{eq:gamma}
\]
where $\pairables=\{\nucG\nucC, \nucC\nucG, \nucA\nucU, \nucU\nucA, \nucG\nucU, \nucU\nucG\}$ is the set of possible base-pairs, $-$ is a gap,  $\gamma'(i,j,S)$ evaluates covariance bonuses and penalties. We follow the 2008 version of RNAalifold~\cite{bernhart+:2008} to use the (symmetric) RIBOSUM matrix $R$ to calculate the covariance, which replaces the Hamming distances $h(s_i, s'_i)$ and $h(s_j, s'_j)$ from the 2002 version of RNAalifold~\cite{hofacker+:2002}:
\[
\hspace{-0.5cm}\gamma'(i,j,S) = \frac{1}{2} \sum_{s, s' \in S,\, s\neq s', (s_i, s_j)\in\pairables, (s'_i, s'_j)\in\pairables}  R(s_i s_j; s'_i s'_j) 
\label{eq:gammaprime}
\]
The basic idea of $\gamma'(i,j,S)$ is to reward {\em compensatory mutations} on column-pair $(i,j)$ across all sequences.
For example, on $(i,j)$ columns, 
if some sequences are AU pairs while others are CG pairs, it is a stronger signal for $(i,j)$ pairing than if all sequences are the same type of pairs.
It is important note that the default version of both \rnaalifold and \linalifold use the 2002 version of $\gamma'(i,j,S)$ but the 2008 version
is substantially more accurate  (which can be invoked by a command-line switch \texttt{-r} in \rnaalifold and \texttt{-r 1} in \linalifold), so we only implemented the 2008 version. All the \rnaalifold and \linalifold results in this paper also used the 2008 version. 
The tunable parameters $\beta$ and $\delta$ are both set to be 1 in \rnaalifold and \linalifold,
but here we tune them using the BL* energy model on the four training families of \rnastralign 
(tRNA, 5S rRNA, tmRNA, Group I Intron) and the best setting is $\beta=1.2$ and $\delta=0.1$.

\subsection*{Correct Calculation of Covariance Bonus $\gamma'(i,j,S)$}
The naive calculation of $\gamma'(i,j,S)$ for each $(i,j)$ column-pair would take $O(k^2)$ time because we need to enumerate all sequence pairs, but RNAalifold employs a clever  method that reduces to $O(k)$ by counting the number of sequences for each type of pair. For example, among $k=8$ sequences, for this $(i, j)$ column-pair, assume we have 5 sequences with CG pairs and 3 with AU pairs,
then we can calculate $\gamma'(i,j,S)$ by aggregating over groups of sequences with the same pair type instead of enumerating all $8\cdot 7/2$ sequence-pairs:
\begin{align*}
\gamma'(i,j,S)=&5\cdot 3 \cdot R(\text{CG};\text{AU})  \\
&+\frac{5\cdot 4}{2} R(\text{CG};\text{CG}) + 
\frac{3\cdot 2}{2} R(\text{AU};\text{AU})
\end{align*}
Or more generally, let $f_{i,j}[t]$ denote the number of sequences with pair type $t$ at $(i,j)$ columns ($t\in \pairables$), then
\hspace{-0.5cm}
\begin{align*}
\gamma'(i,j,S)= &\sum_{t,t'\in \pairables, t\neq t'} f_{i,j}[t]\cdot f_{i,j}[t']\cdot R(t;t') \\
&+ \sum_{t\in \pairables} \binom{f_{i,j}[t]}{2}\cdot R(t;t)
\end{align*}
The first term calculates the contribution from compensatory mutations (different pair types $t$ and $t'$) and the second term calculates the contribution from the same pair type $t$.

\smallskip

However, it is worth noting that both RNAalifold and LinAliFold calculated this term incorrectly. 

\begin{enumerate}[leftmargin=20pt]
    \setlength{\itemsep}{0pt}    
    \item RNAalifold (and LinAliFold by inheritance) uses an oversimplified formula:
\[
\gamma'(i,j,S)= \sum_{t,t'\in \pairables} f_{i,j}[t]\cdot f_{i,j}[t']\cdot R(t;t')
\]
which incorrectly handles the score contributions for sequences that have the same base pair type $t$ at positions $(i, j)$. 
The correct method (as shown above) should multiply the RIBOSUM score $R(t; t)$ by $\binom{f_{i,j}[t]}{2}$ which reflects the correct number of pairwise comparisons without repetition among sequences while RNAalifold and LinAliFold erroneously calculates this as $(f_{i,j}[t])^2 \cdot R(t; t)$. 
    Note that this miscalculation also includes comparisons of each sequence with itself, 
    i.e., pairs $(s_i s_j; s'_i s'_j)$ where $s = s'$, which should not contribute to the score since they do not provide information about covariation.
    This overcounting inflates the conservation score, leading to potentially incorrect results in RNA structural predictions.
   
    \item Another computational error in RNAalifold is the normalization of the $\gamma'(i,j,S)$ term. The correct approach should normalize this term by $k^2$, reflecting the total number of pairwise sequence comparisons.
    However, RNAalifold incorrectly normalizes this term by just $k$. This insufficient normalization leads to amplified contributions from sequence pairs, therefore distorting the score. LinAliFold corrected this issue and computes the normalization correctly.
\end{enumerate}

\subsection{Partition Function Mode}

The consensus partition function $Q(S)$ over a set $S$ of aligned sequences is:
\[
Q(S) = \sum_y \exp({-\score(S, y) / RT})
\]
where $R$ is the molar gas constant and $T$ is the absolute temperature.
The Boltzmann probability of a consensus structure $y$ is then:
\[
p(y \mid S) = \frac{\exp({-\score(S, y) / RT})}{Q(S)}
\]
and the consensus (marginal) base-pairing probability that column $i$ is paired with column $j$ is:
\[
p_{ij}(S) = \sum_{y: (i,j)\in y} p(y \mid S)
\]
When projecting this consensus base-pairing matrix down to each individual sequence,
we delete columns and rows that are dashes ($-$) in that sequence,
as well as $p_{ij}$ entries that correspond to non-pairable bases in that sequence.

\subsection*{LazyOutside Algorithm}
\label{sec:lazyoutside}
By default, the inside-outside algorithm \cite{baker:1979} is used 
to calculate the marginal base-pairing probabilities,
where the McCaskill algorithm \cite{mccaskill:1990} is a special case.
Conventionally, the outside phase is considered a mirror image of the inside phase, with similar or slower runtimes, which means inside-outside is (at least) twice as slow as their inside-only or MFE. 
We employ our unpublished technique of LazyOutside \cite{huang+:2024} 
which is a lazy (on-demand) algorithm 
that only visits high-probability states and ignores the low-probability ones. 
Basically, let us denote $\alpha(v)$ to be the inside partition function for node $v$ (e.g., $\text{P}_{5,10}$) and $\beta(v)$ to be the outside partition function, then we prune nodes $v$ if its marginal probability falls under a threshold $\theta$:
\[
\alpha(v) \cdot \beta(v) / Q(S) < \theta
\]
and we use the default $\theta=5\times 10^{-5}$.
This pruning was also used in natural language parsing (``relative useless pruning'') and machine learning (``max-marginals'') \cite{huang:2008}.
As a result, it only visits a tiny fraction (often as small as 1\%) of the states visited in the inside phase, which implies up to 100× speedup of the outside phase, making inside-outside almost as fast as the inside phase alone.


\subsection*{Structural Distance and Ensemble Defect}

We employ structural distance and ensemble defect~\cite{Zadeh+:2010} as two key metrics to evaluate the prediction accuracy of our tool. Structural distance is basically a structured version of Hamming distance between two structures, while ensemble defect is the expectation of structural distance in the Boltzmann ensemble.

More formally, let \vecx be an RNA sequence and \vecy and \vecystar be two secondary structures of \vecx. The structural distance between \vecy and \vecystar quantifies the structural discrepancies between them,
specifically in terms of mismatched base pairs and unpaired nucleotides, calculated using the following formula:
\begin{align*}
    d(\vecy, \vecystar)\  =\  & |\vecx| - 2 | \text{pairs}(\vecy) \cap \text{pairs}(\vecystar)| \\
    & - |\text{unpaired}(\vecy) \cap \text{unpaired}(\vecystar)|
\end{align*}

The ensemble defect is used to quantify the deviation of an RNA ensemble from a target structure \vecystar, which is the expectation of structural distance to \vecystar over the Boltzmann ensemble:
\[
 \Phi(\vecx, \vecystar) = \textstyle\E_{\vecy\, \sim\, p(\;\cdot
 \ \mid\ \vecx)} \ [d(\vecy, \vecystar)]
\]
On the surface, this definition seems to range over all possible structures in the expectation,
but we can use dynamic programming to factor this computation 
to the expected number of incorrectly predicted nucleotides over the whole ensemble at equilibrium:
\[\hspace{-0.4cm}
\Phi(\vecx, \vecystar) = |\vecx| - 2 \hspace{-.2cm}\sum_{(i,j) \in \text{pairs}(\vecystar)} p_{i,j}(\vecx) - \hspace{-.2cm}\sum_{j \in \text{unpaired} (\vecystar)} q_j(\vecx)
\]
where \( p_{i,j}(\vecx) \) is the probability of nucleotide \( i \) pairing with nucleotide \( j \), and \( q_j(\vecx) \) is the probability of nucleotide \( j \) being unpaired, defined as \( q_j = 1 - \sum p_{i,j} \).

\subsection*{RNAstralign Datasets}
We use a procedure similar to LinearTurboFold \cite{li+:2021} to sample homologs from the RNAstralign dataset.
Four families (Group I Intron, tmRNA, tRNA, and 5S rRNA) are used for tuning and another four families (SRP, RNaseP, telomerase, and 16S rRNA) are used for testing.
For Group I Intron, 5S rRNA, SRP, RNaseP, and 16S rRNA,
there are multiple subfamilies within each family, so we chose one specific subfamily for these five families (see Tab.~\ref{tab:RNAstralign}).
For 16S rRNA, we also made sure that only full-length sequences (rather than subdomains) are included.
For each (sub)family, we drew 10 samples, each with $k=30$ homologs and align them by MAFFT \verb|--auto|. 
Tab.~\ref{tab:RNAstralign} also presents the average sequence length and average sequence identity of the MSAs in each family. 
These values are specifically calculated for the dataset of 10 samples (with $k=30$ homologs per sample), with sequence identity determined from the alignments performed by MAFFT \verb|--auto|. 
This data is used for the evaluations shown in Fig.~\ref{fig:accuracy} and Fig.~\ref{fig:si-accuracy-10}.
We included all these samples in our GitHub.

\begin{table}[t]
\caption{RNAstralign benchmark dataset. 
These values are specifically calculated for the dataset of 10 samples (with $k=30$ homologs per sample), with sequence identity determined from the alignments performed by MAFFT \texttt{--auto}.
This data is used for the evaluations shown in Fig.~\ref{fig:accuracy} and Fig.~\ref{fig:si-accuracy-10}.}
\resizebox{.48\textwidth}{!}{
\begin{tabular}{ccrr}
family & subfamily & avg.~seq.~len. & avg.~seq.~identity\\
\hline
Group 1 & IC1 & 428.5 & 0.36 \\
tmRNA & \-- & 367.4 & 0.42 \\
tRNA & \-- & 77.1 & 0.48 \\
5S rRNA & Bacteria & 116.2 & 0.62 \\
\hline
SRP & Protozoan & 285.8 & 0.36 \\
RNaseP & A bacterial & 360.0 & 0.45 \\
telomerase & \-- & 444.9 & 0.50 \\
16S rRNA & Alphaproteobacteria & 1419.2 & 0.85
\end{tabular}
}
\label{tab:RNAstralign}
\end{table}

\subsection*{\sarscovtwo and \sarsr Datasets}
We prepared a dataset to draw representative samples of  diverse \sarscovtwo and \sarsr genomes.
Based on the genomes from GISAID~\cite{elbe+:2017} (downloaded on 4 April 2022) and NCBI (\url{www.ncbi.nlm.nih.gov}; genomes submitted from 1998 to 2019), we first filtered out low-quality genomes (i.e., those with unknown characters or are shorter than $28,000\nts$).
After preprocessing, we obtained two datasets with $\sim40,000$ \sarscovtwo (including Alpha, Beta, Delta, and Omicron variants) and $600$ \sarsr genomes, respectively.
Following \linearturbofold, we used a sampling algorithm to choose 60\%  diverse \sarscovtwo genomes and 40\%  diverse \sarsr genomes (see Tab.~\ref{tab:covid}).
Unlike LinearTurboFold\cite{li+:2021}, we did not use a greedy algorithm to choose the most diverse genomes one by one,
but only randomly sample for each category (Alpha, Beta, Delta, Omicron, SARS-related).
We included all the COVID samples in our GitHub.

\begin{table}
\caption{SARS-CoV-2 and SARS-related  datasets. Ref is the SARS-CoV-2 reference sequence, Alpha--Delta are the SARS-CoV-2 variants, and SARSr are SARS-related genomes.}
\centering
\resizebox{0.48\textwidth}{!}{
\begin{tabular}{r|rrrrr|r}
$k$ & Ref & Alpha & Beta & Delta & Omicron & SARSr \\
\hline
10 & 1 & 2 & 2 & 1 & 1 & 3 \\
30 & 1 & 4 & 5 & 4 & 4 & 12 \\
50 & 1 & 7 & 8 & 7 & 7 & 20 \\
100 & 1 &14 & 15 & 15 & 15 & 40 \\
200 & 1 & 33 & 33 & 33 & 20 & 80 \\
300 & 1 &59 & 60 & 40 & 20 & 120 \\
400 & 1 &79 & 80 & 60 & 20 & 160 \\
\end{tabular}
}
\label{tab:covid}
\end{table}

\subsection*{Hybrid Reference Structure Construction in the 5' and 3' UTR regions of \sarscovtwo}
To get the hybrid reference structure in the UTR regions (Fig.~\ref{fig:hybrid}H), we combined the experimentally guided structures from Huston et al.~\cite{huston+:2021} and the experimentally determined end-to-end pairs (Arch3, ranges from (60,29868) to (80,29847)) from Ziv et al.~\cite[Fig.~3]{ziv+:2020} by the following steps:
\begin{enumerate}[leftmargin=20pt]
\setlength{\itemsep}{0pt}
\item Get (local) structures in 5' and 3' UTR regions from Huston et al.~(the 5' UTR ranges from 1 to 400 and the 3' UTR from 29543 to 29876 on the reference sequence).
\item Remove (local) pairs $(i,j)$ from the structures if $i$ or $j$ is in the global Arch3 pairs (e.g., SL3 from~Huston et al.~\cite[Fig.~2]{huston+:2021} is removed). These local pairs were predicted by the local folding software which can only predict pairs within a local window.
\item Combine the modified structures and the end-to-end Arch3 pairs from Ziv et al.
\end{enumerate} 
See Fig.~\ref{fig:hybrid}H for details; we also released its dot-bracket format on our Github. This hybrid structure is used for evaluating prediction qualities in Figs.~\ref{fig:covid} \&~\ref{fig:covid-si}.


\subsection*{Software and Computing Environment}
We use the following software:
\vspace{-0.1cm}
\begin{itemize}[leftmargin=10pt]
\setlength{\itemsep}{-3pt}
\item \rnaalifold (Vienna RNAfold 2.4.16) (\verb|-r| mode)\\
\url{https://www.tbi.univie.ac.at/RNA/}
\item MAFFT 7.490 (always with \verb|--auto| mode)\\
\url{https://mafft.cbrc.jp/alignment/software/}
\item \linearalifold (\verb|-r 1|) 
\url{https://github.com/fukunagatsu/LinAliFold-CentroidLinAliFold}
\item \linearturbofold\\
\url{https://github.com/LinearFold/LinearTurboFold}
\end{itemize}

We benchmarked these tools on a Linux machine with 2 Intel Xeon E5-2660 v3 CPUs (2.60 GHz) and 377 GB memory, and used \verb|gcc| (Ubuntu 9.3.0-17) to compile.


\section*{Code and Data Availability}
Our code and data are released on GitHub:\\
\url{http://github.com/LinearFold/LinearAlifold}.


\noindent 
Server at: \url{http://linearfold.org/linear-alifold}.

\vspace{-.2cm}
\section*{Conflict of Interest}
The authors declare no conflict of interest.

\vspace{-.2cm}
\section*{Author Contributions}

L.H.~conceived the idea and directed the project.
L.Z.~designed the main algorithm and implemented the initial version;
A.M.~reimplemented the whole system in much higher quality,
implemented LazyOutside, ThreshKnot, and alifold-aware stochastic sampling, added BL* energy model, performed parameter tuning, 
and conducted thorough evaluations against \linalifold, \rnaalifold, and \linearturbofold\ on SARS-CoV-2 and \rnastralign.
A.M.~also built the web server.
M.G.~made the visualizations of COVID structures and stochastic sampling.
N.D.~contributed to the evaluations on SARS-CoV-2.
S.L.~contributed to the evaluations on both \rnastralign\ and SARS-CoV-2, as well as the comparison to \linearturbofold.
H.Z.~contributed to the algorithm design.
D.H.M.~guided the evaluations.
L.H., L.Z., A.M., S.L., H.Z., and D.H.M.~wrote the manuscript.

\section*{Acknowledgments}

This work was supported in part by National Institutes of Health Grant R35GM145283 (to D.H.M.) and National Science Foundation Grants 2009071 (to L.H.)
and 2330737 (to L.H.~and D.H.M.).
We thank the reviewers for insightful suggestions which greatly improved the quality of our work.
We also thank Tsukasa Fukunaga and Michiaki Hamad (authors of LinAlifold) for citing our preprint version.



\bibliographystyle{unsrt}

\bibliography{refs}

\newpage


\captionsetup[figure]{font={stretch=1}} 
\renewcommand{\thefigure}{S\arabic{figure}}
\renewcommand{\thetable}{S\arabic{table}}

\renewcommand\thesection{\S\arabic{section}}


\setcounter{figure}{0} 

\begin{figure*}[!t]
\centering
{\Large \bf Supplementary Figures}
\end{figure*}

\begin{figure*}[!t]
    \centering
    \begin{tabular}{c}
    {\panel{A}}\\[1mm]
    \includegraphics[width=.88\linewidth]{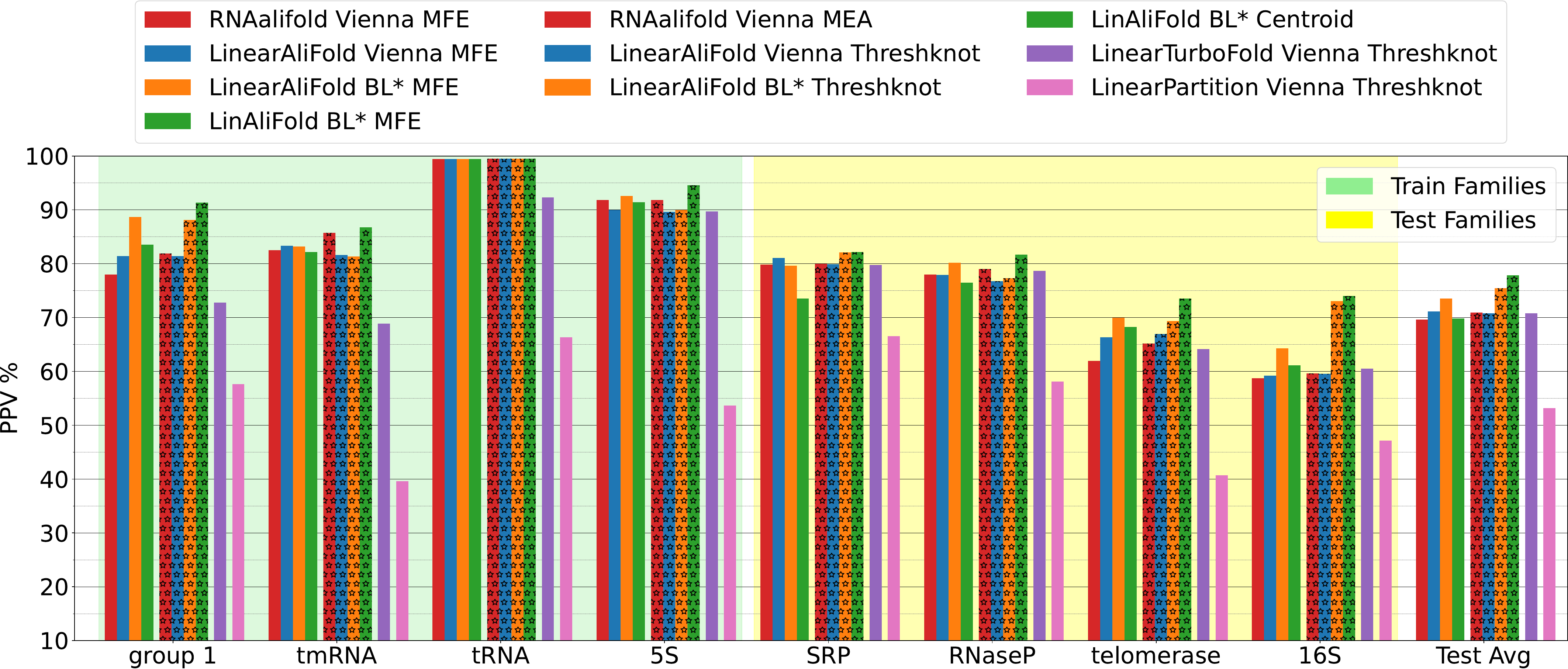}\\
    [2.5mm]
    {\panel{B}}\\[-5mm]\\
    \includegraphics[width=.9\linewidth]{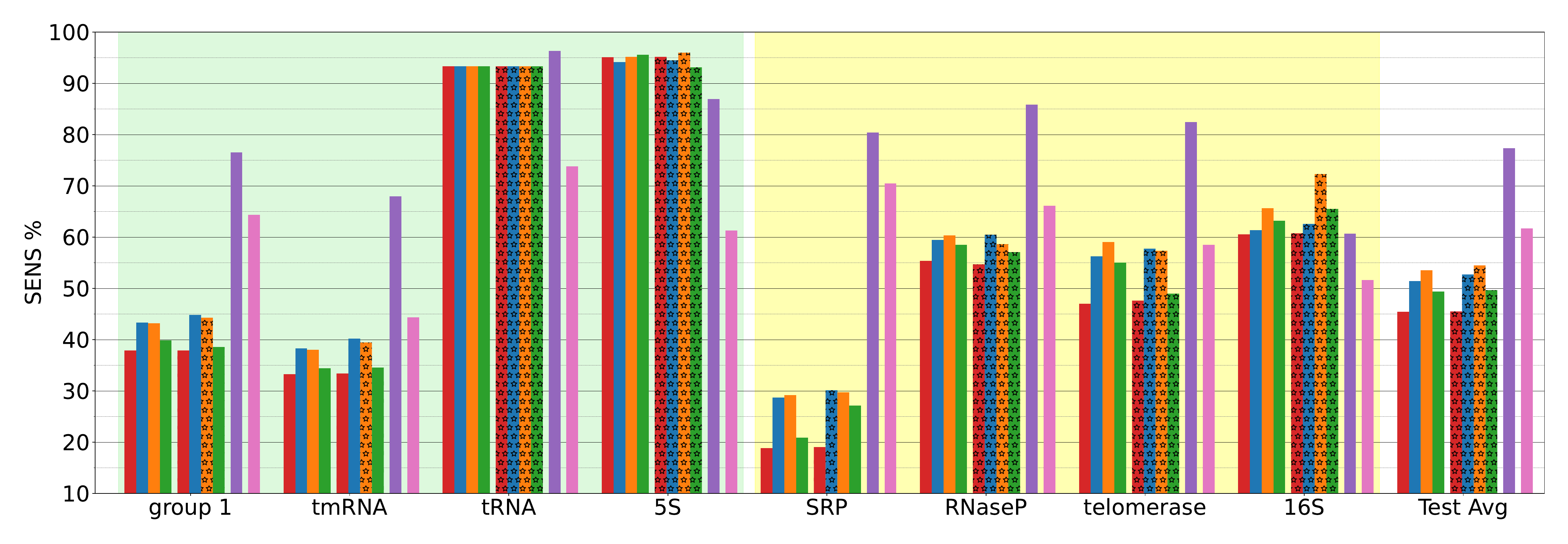}\\
    {\panel{C}}\\[-5mm]\\
    \includegraphics[width=.9\linewidth]{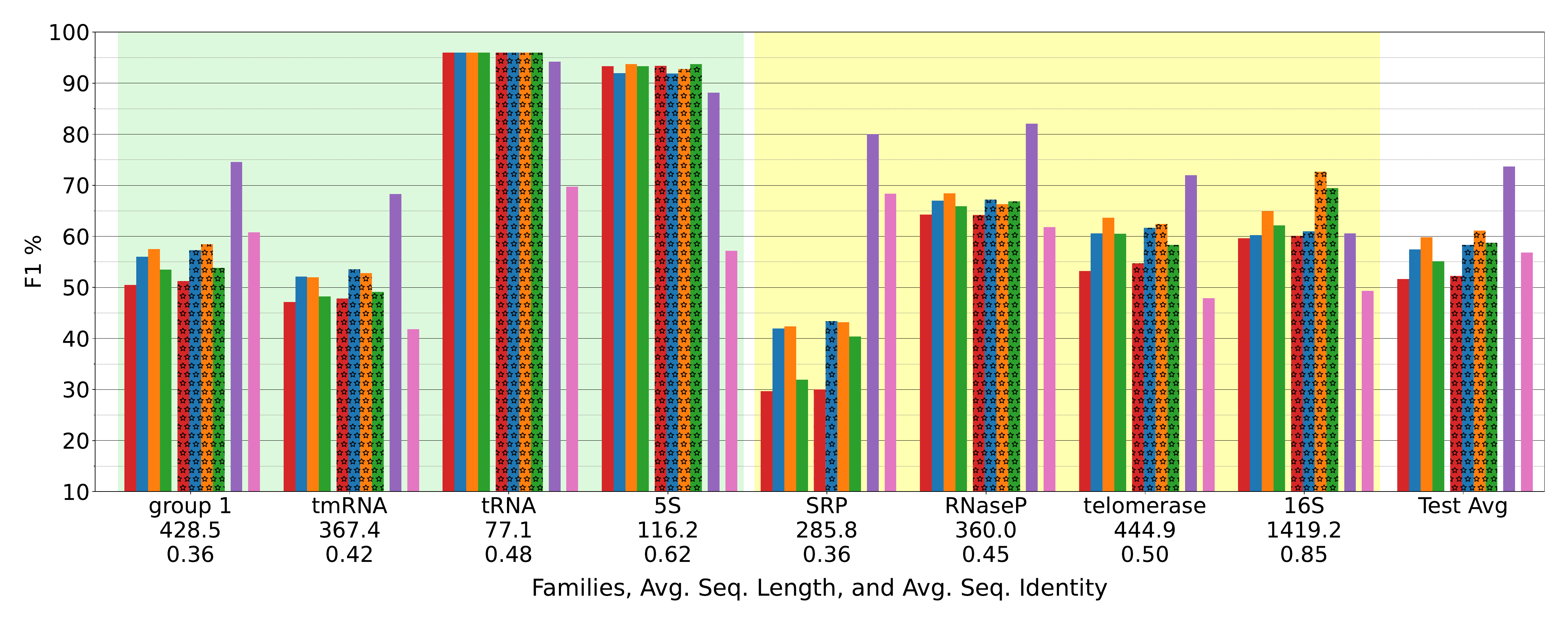} 
    \end{tabular}
    \caption{Accuracy comparisons on the RNAstralign dataset, similar to Fig.~\ref{fig:accuracy} but including more systems. Each family has 10 samples, and each sample is an MSA with $k=30$ homologs. 
    Align-then-fold systems (RNAalifold, LinearAlifold, and LinAliFold) tend to be inaccurate for low sequence indentity families (e.g., SRP and group 1) and tend to be more accurate for high sequence identity 
    families (e.g., 16S rRNA). Refer to Fig.~\ref{fig:si-accuracy-20} for a similar figure with 20 samples per family.
    \label{fig:si-accuracy-10}
    }
\end{figure*}

\begin{figure*}[!t]
    \centering
    \begin{tabular}{c}
    {\panel{A}}\\[1mm]
    \includegraphics[width=.88\linewidth]{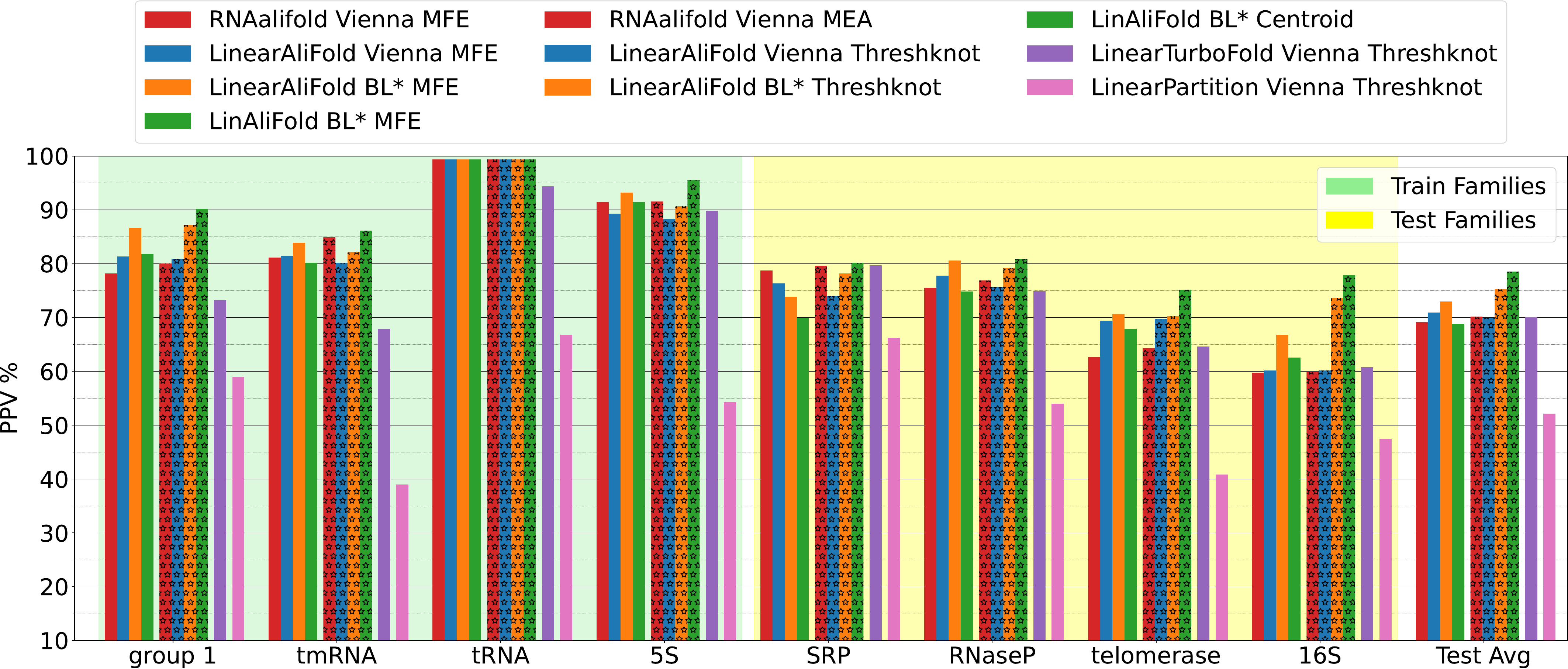}\\
    [2.5mm]
    {\panel{B}}\\[-5mm]\\
    \includegraphics[width=.9\linewidth]{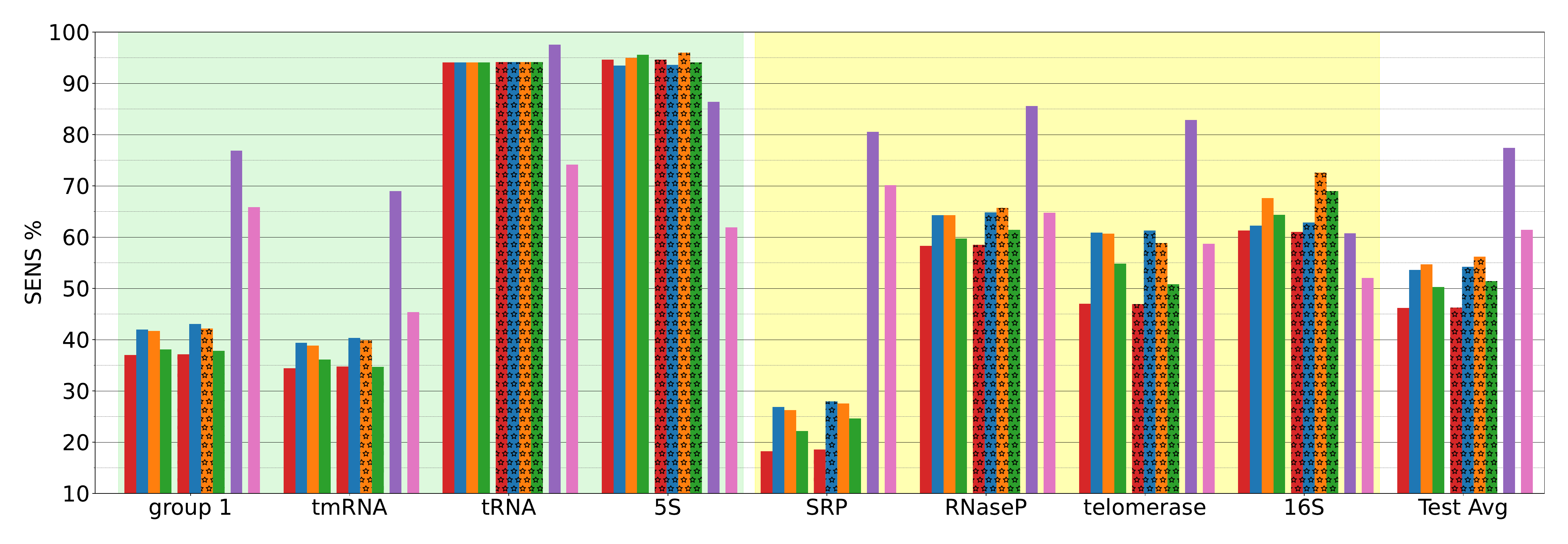}\\
    {\panel{C}}\\[-5mm]\\
    \includegraphics[width=.9\linewidth]{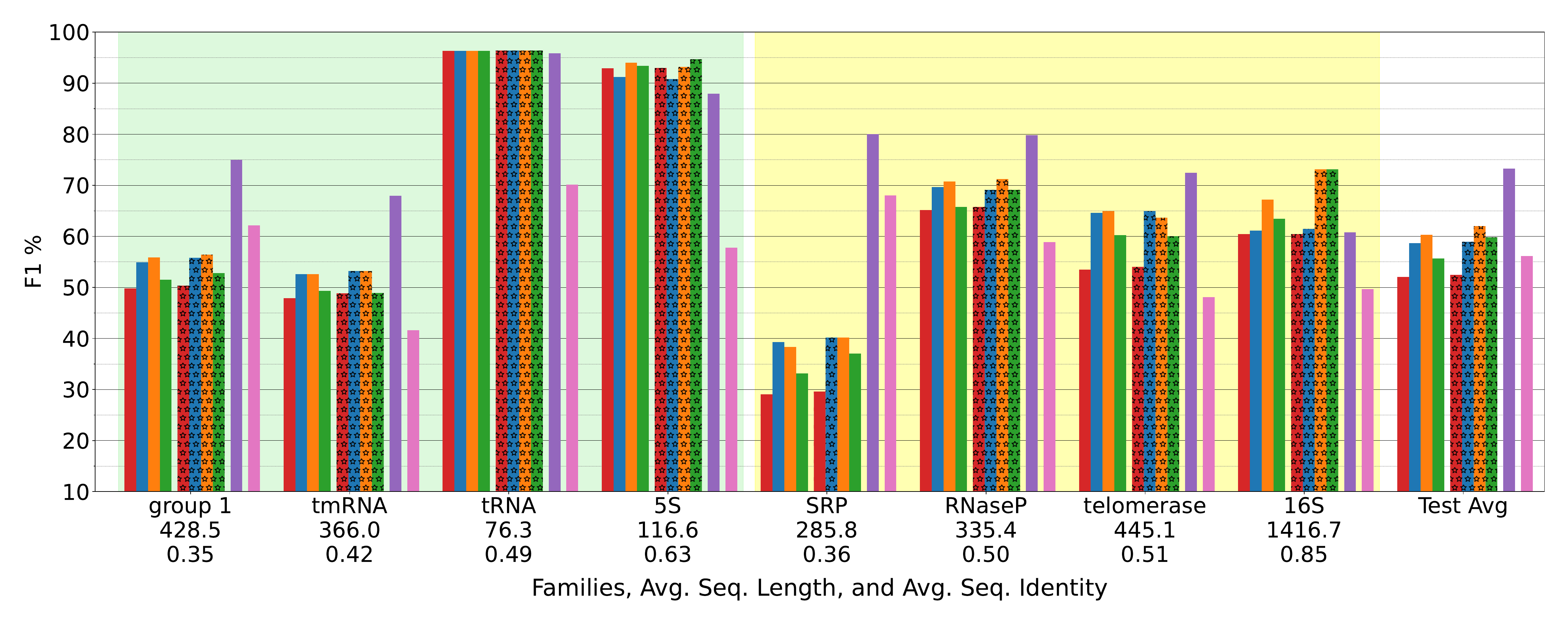} 
    \end{tabular}
    \caption{Accuracy comparisons on the RNAstralign dataset. Each family has 20 samples, and each sample is an MSA with $k=30$ homologs. 
    Align-then-fold systems (RNAalifold, LinearAlifold, and LinAliFold) tend to be inaccurate for low sequence indentity families (e.g., SRP and group 1) and tend to be more accurate for high sequence identity 
    families (e.g., 16S rRNA). Refer to Fig.~\ref{fig:si-accuracy-10} for a similar figure with 10 samples per family.
    \label{fig:si-accuracy-20}
    }
\end{figure*}

\begin{figure*}[!t]
\hspace{-0.6cm}
\begin{tabular}{ccc}

& \multicolumn{1}{c}{\large{Vienna Energy Model (A, C, E)}} & \multicolumn{1}{c}{\large{BL* Energy Model (B, D, F)}} \\[5pt]

& \multicolumn{1}{c}{\textbf{A}} & \multicolumn{1}{c}{\textbf{B}} \\
& \includegraphics[width=.48\linewidth]{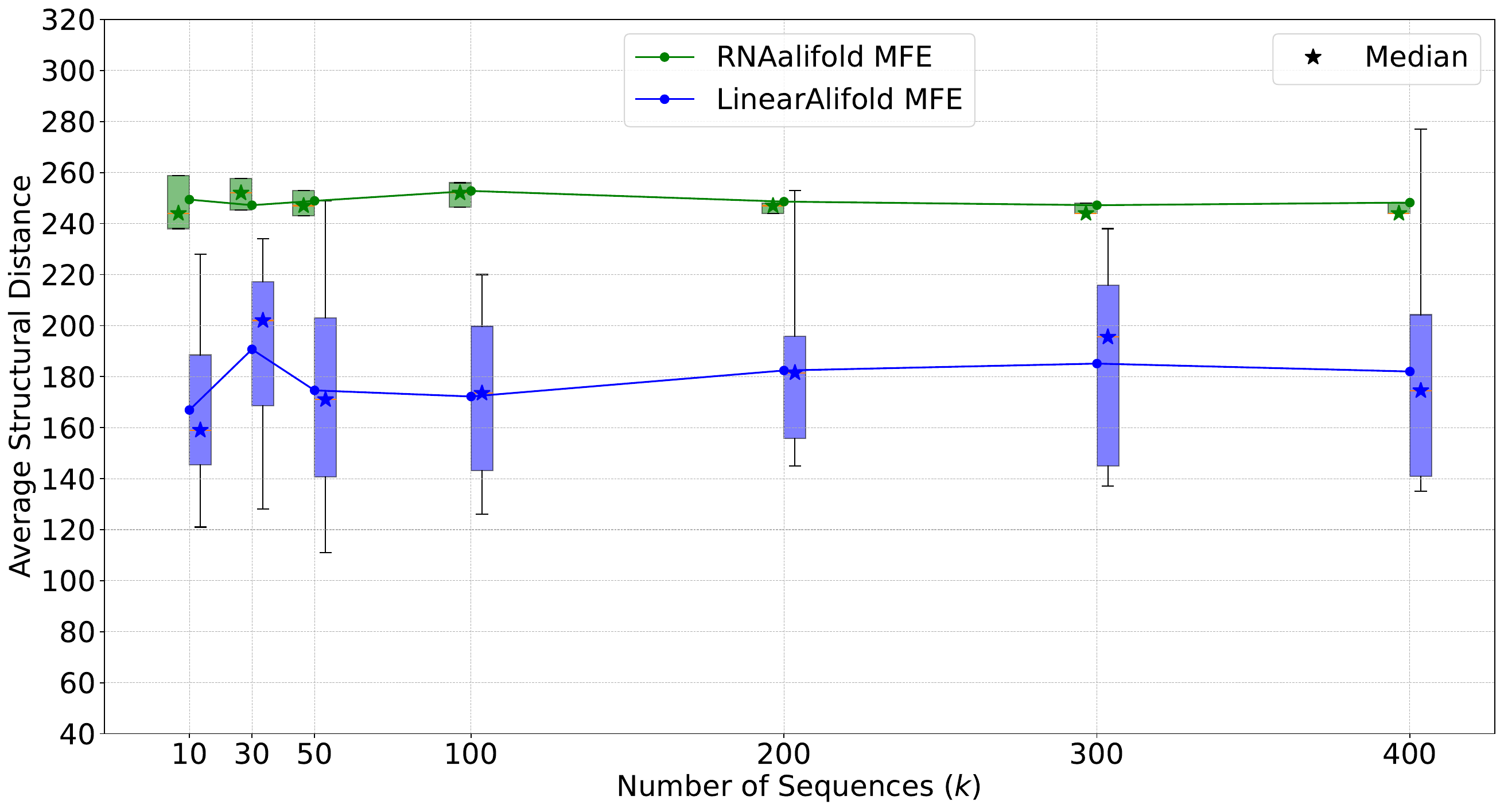}
& \includegraphics[width=.48\linewidth]{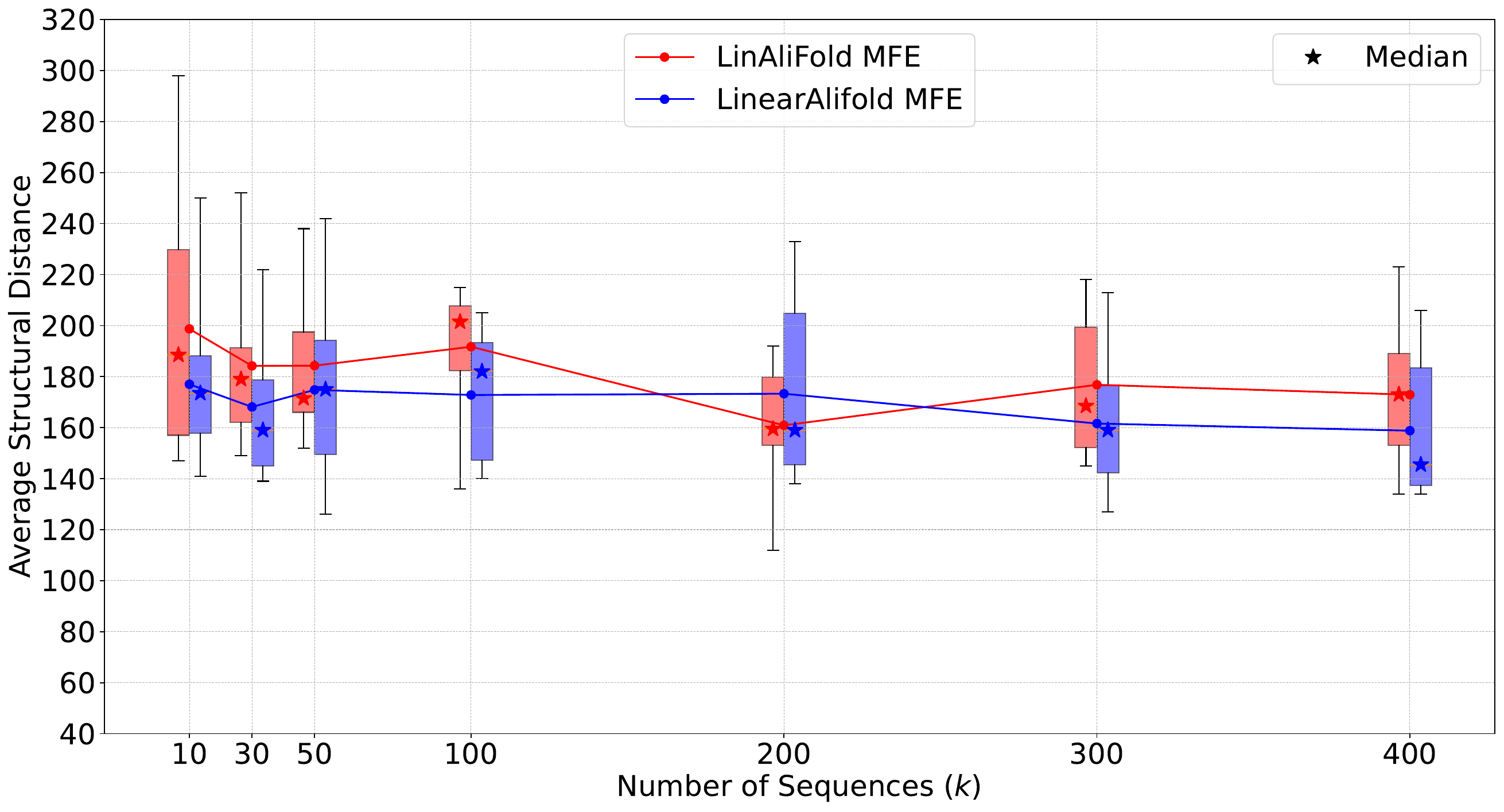} \\

& \multicolumn{1}{c}{\textbf{C}} & \multicolumn{1}{c}{\textbf{D}} \\
& \includegraphics[width=.48\linewidth]{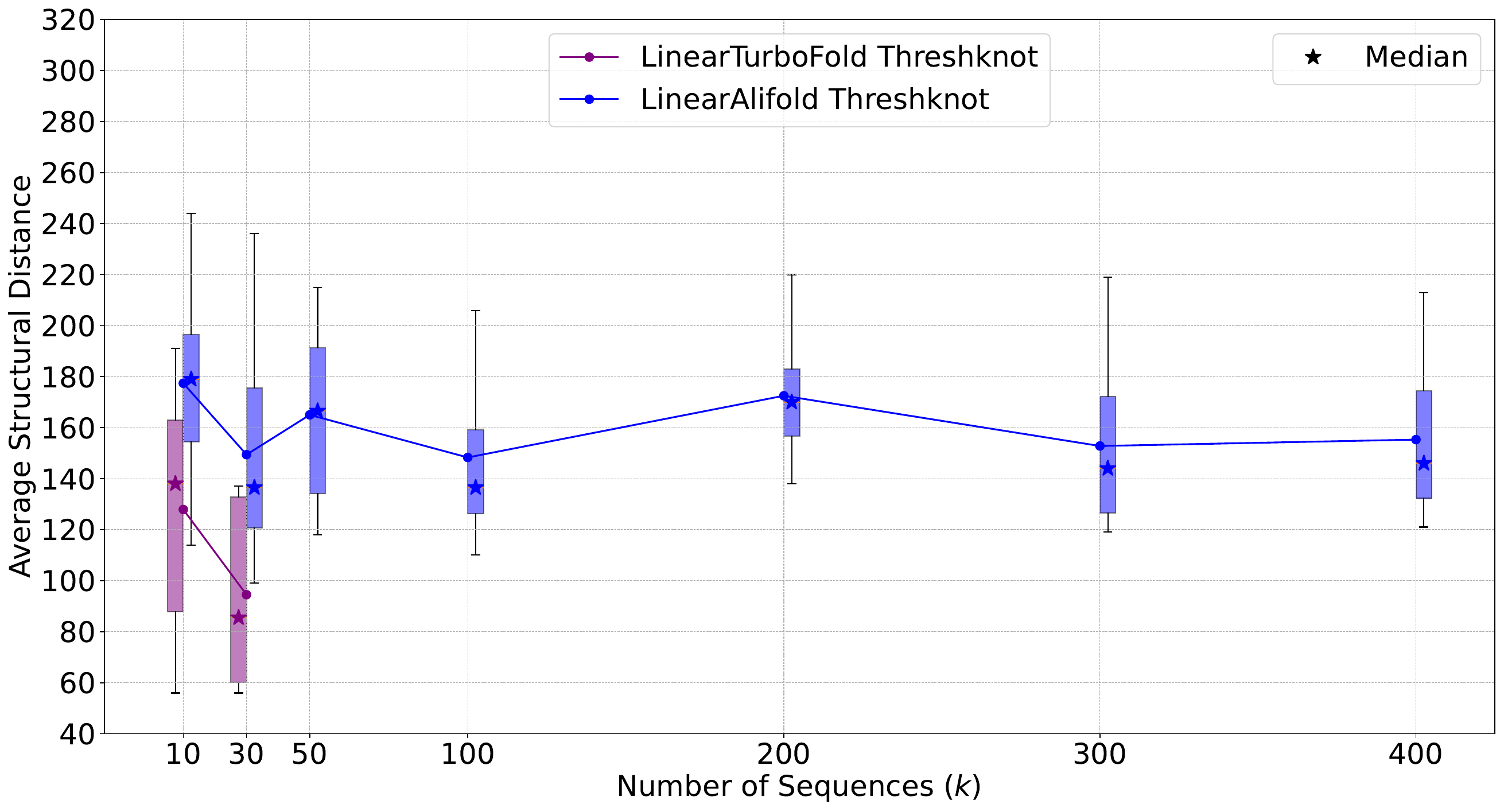}
& \includegraphics[width=.48\linewidth]{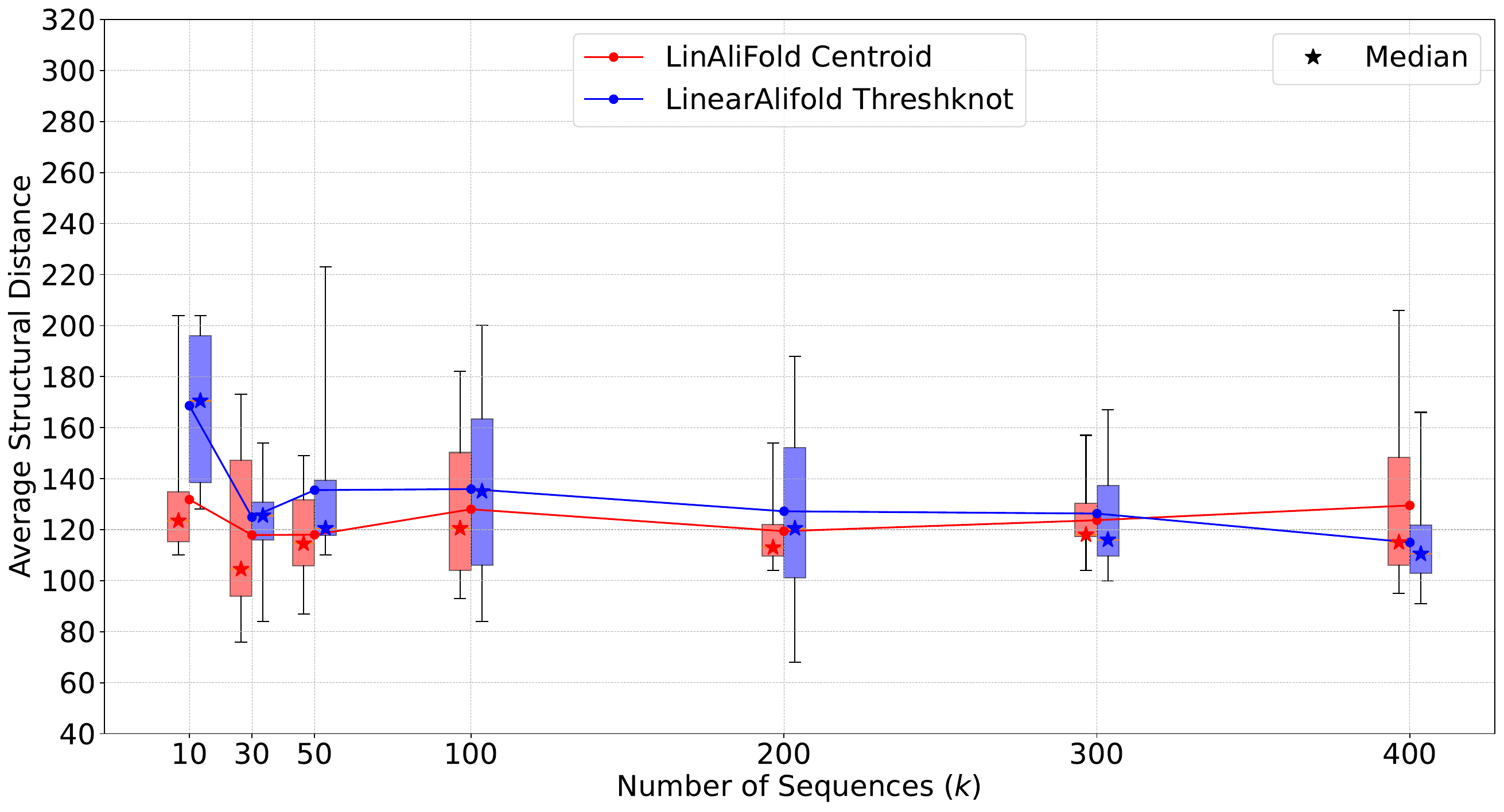} \\

& \multicolumn{1}{c}{\textbf{E}} & \multicolumn{1}{c}{\textbf{F}} \\
& \includegraphics[width=.48\linewidth]{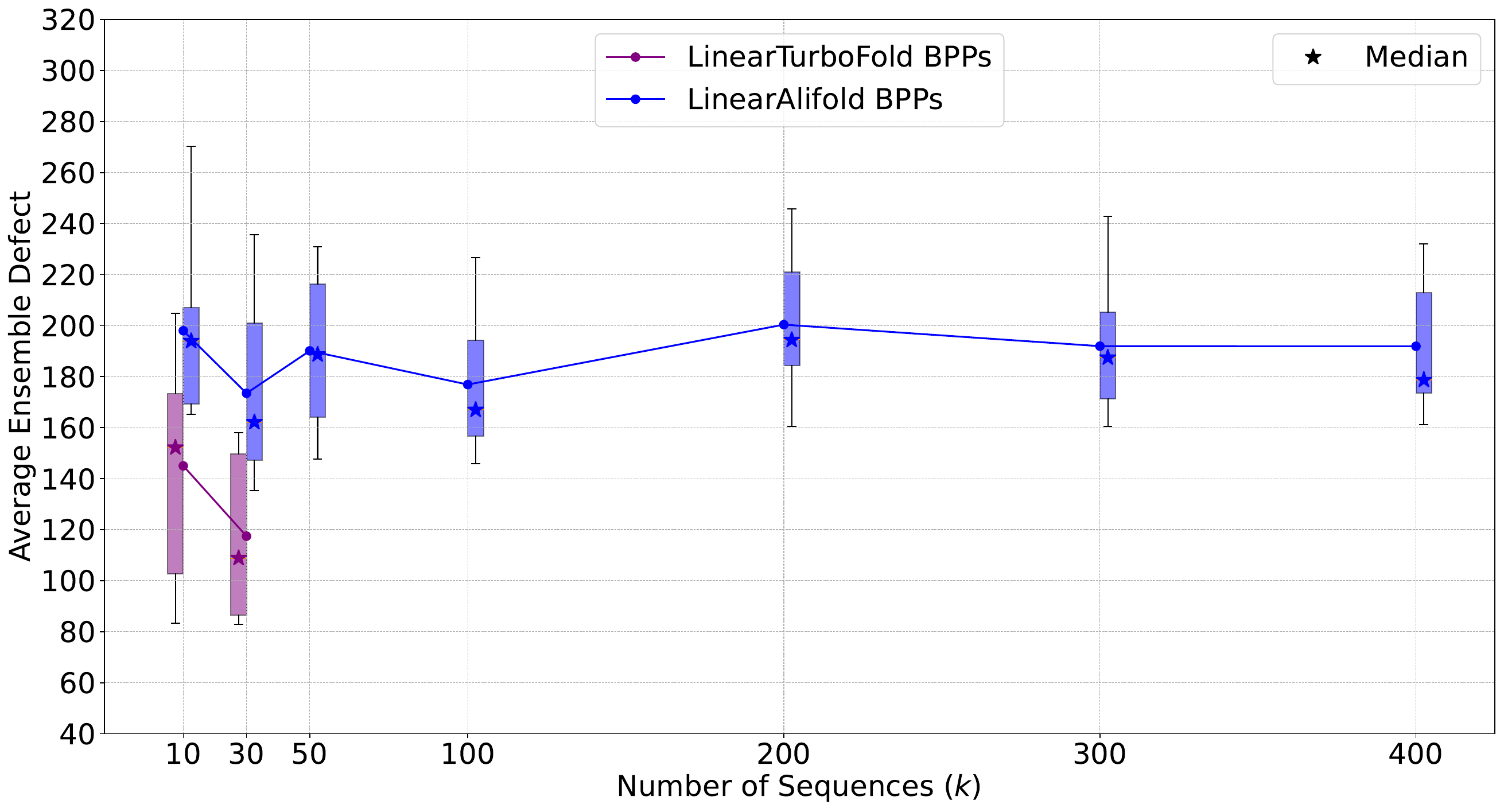}
& \includegraphics[width=.48\linewidth]{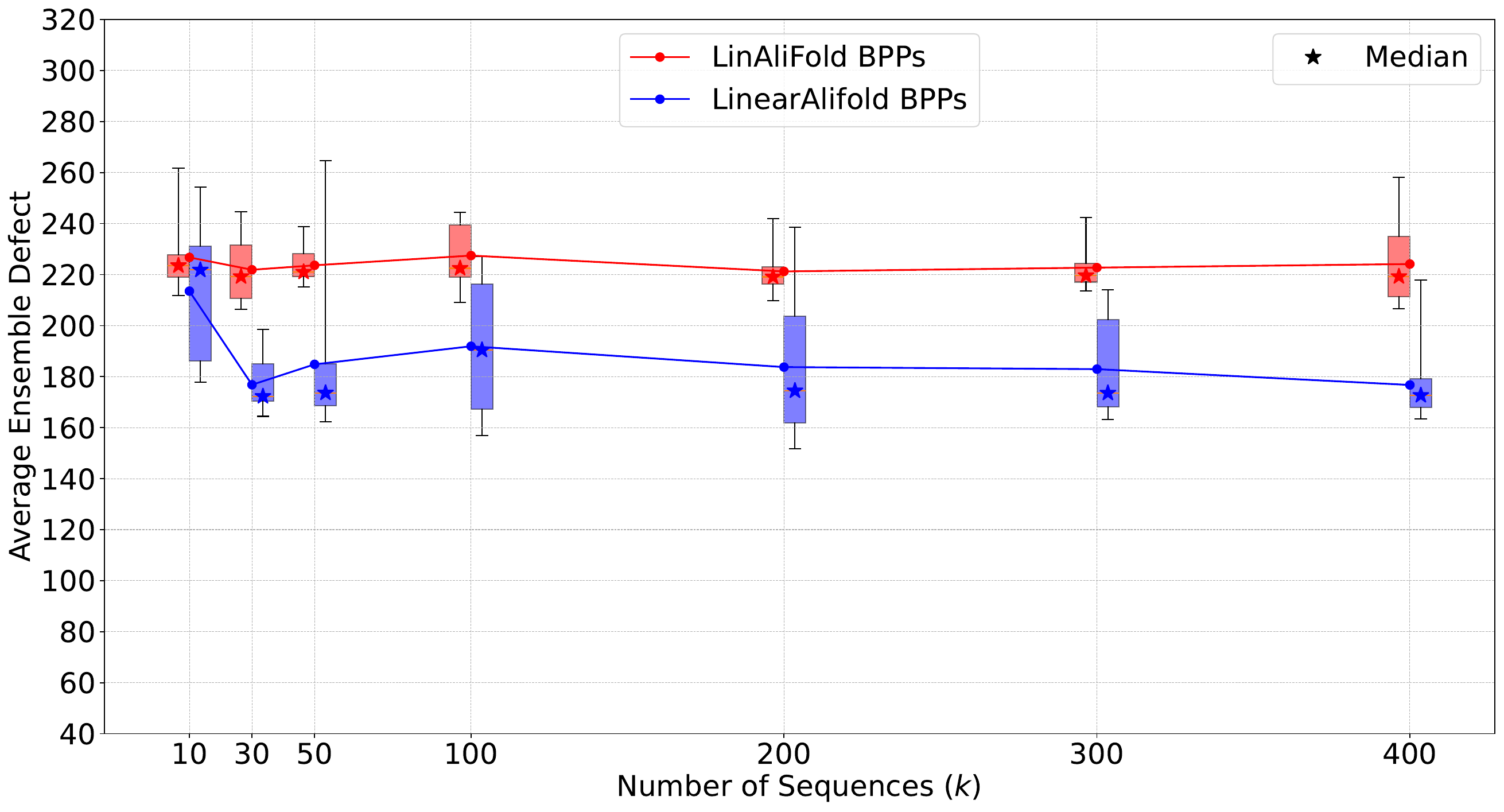} \\

\end{tabular}
\caption{Box plot of structural distance and ensemble defect (of 5' and 3' UTRs) against the number of sequences ($k$) for different energy models and methods, with the structural distance evaluated using the hybrid COVID structure (Fig.~\ref{fig:hybrid}H; Huston et al. + Ziv et al.) as the reference (see Fig.~\ref{fig:huston-only} for a similar figure with the Huston et al.~structure as reference).
For each $k$, we have 10 samples, so the boxes show the 25-75 percentiles, and the whiskers represent the full range of data (1-100 percentile). 
The curves show the mean values over 10 samples for each $k$, and the stars denote the medians. See Fig.~\ref{fig:covid} for another version where the x-axes are time instead of $k$.
A--B: MFE prediction. C--D: partition-based structure prediction.
E--F: ensemble quality.}
\label{fig:covid-si}
\end{figure*}

\begin{figure*}[!t]
    \hspace{-0.4cm}
    \begin{tabular}{ccc}
    
    & \multicolumn{1}{c}{\large{Vienna Energy Model (A, C, E)}} & \multicolumn{1}{c}{\large{BL* Energy Model (B, D, F)}} \\[5pt]
    
    & \multicolumn{1}{c}{\textbf{A}} & \multicolumn{1}{c}{\textbf{B}} \\
    & \includegraphics[width=.48\linewidth]{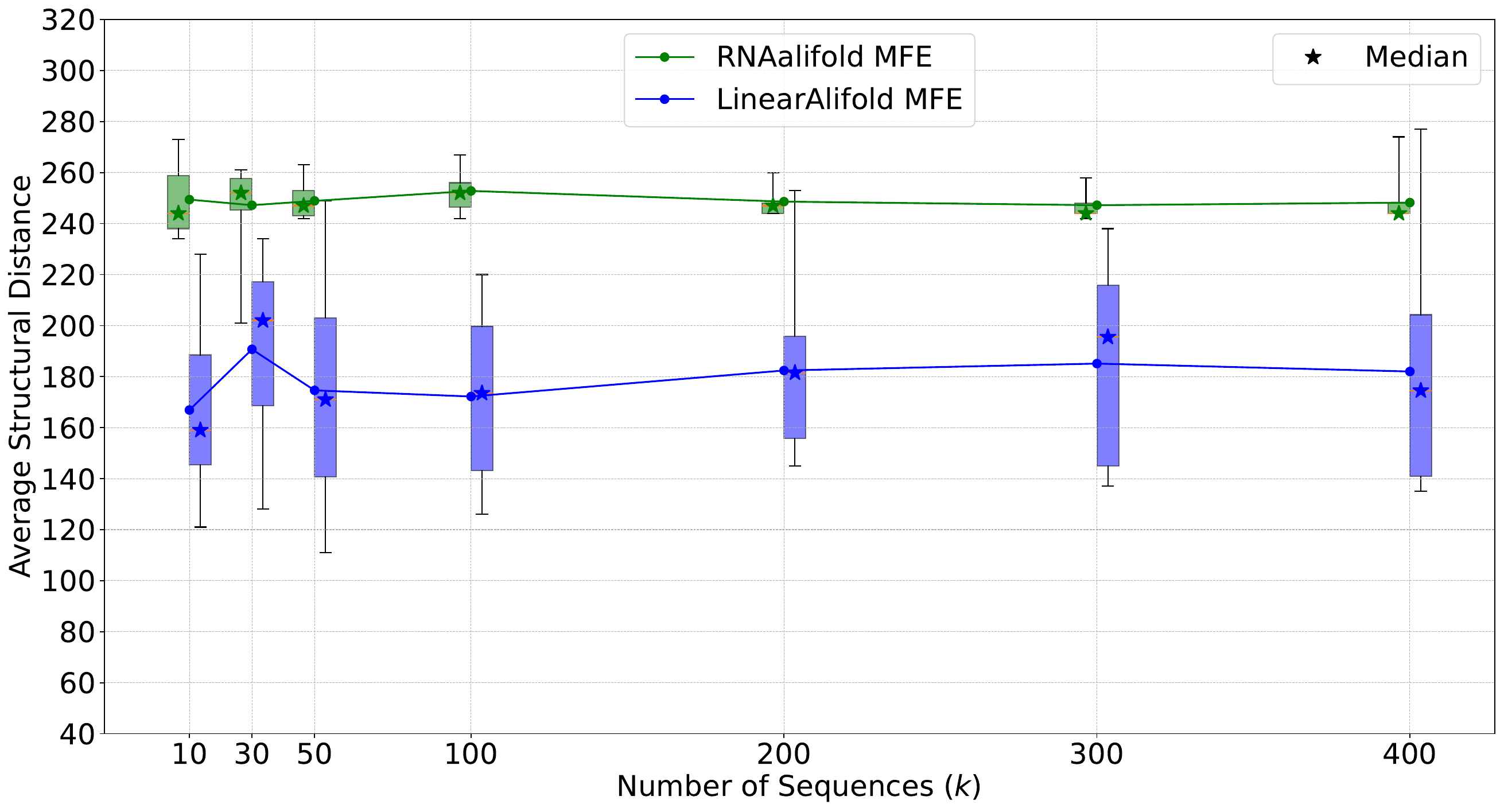} 
    & \includegraphics[width=.48\linewidth]{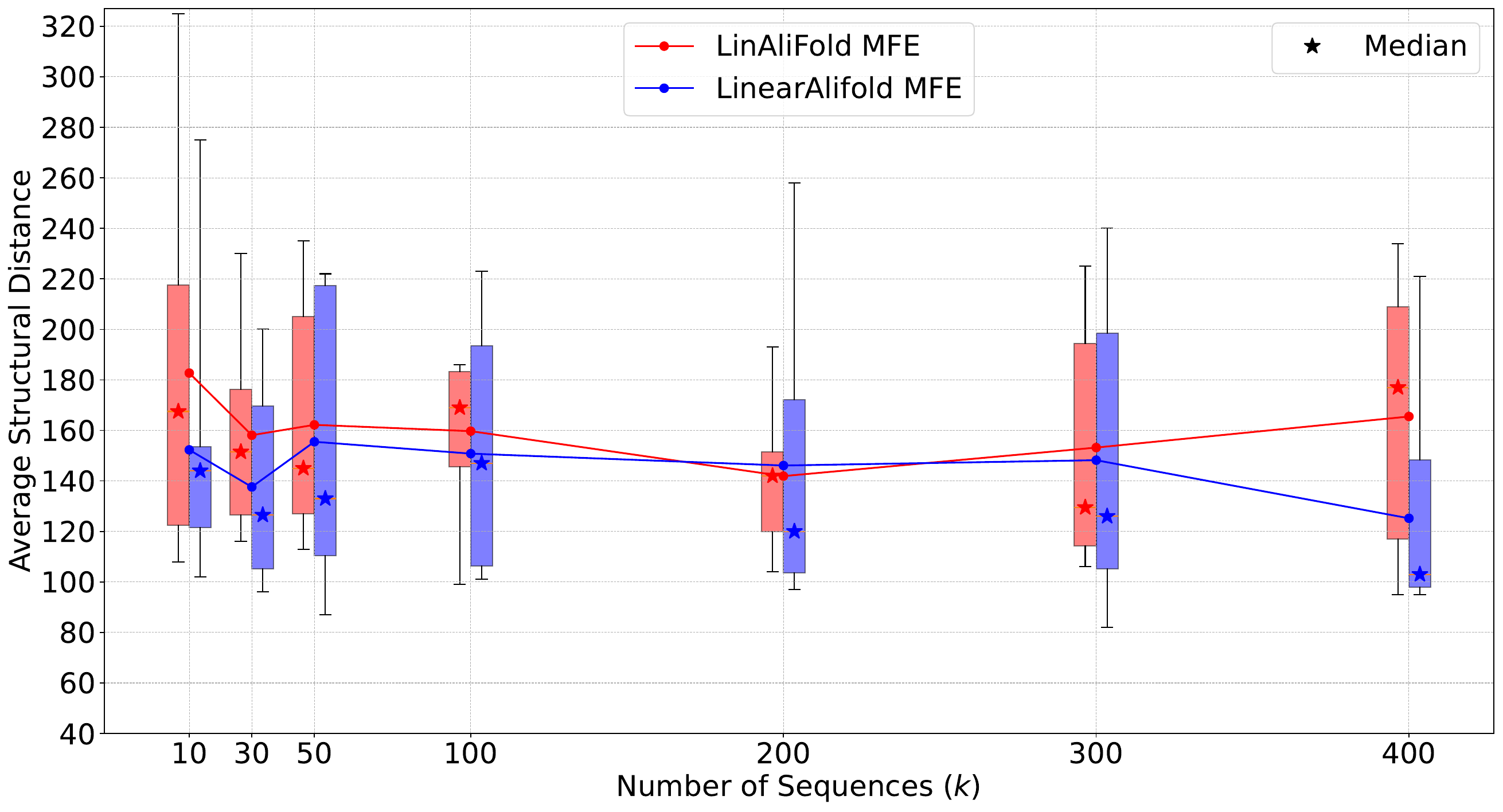} \\
    
    & \multicolumn{1}{c}{\textbf{C}} & \multicolumn{1}{c}{\textbf{D}} \\
    & \includegraphics[width=.48\linewidth]{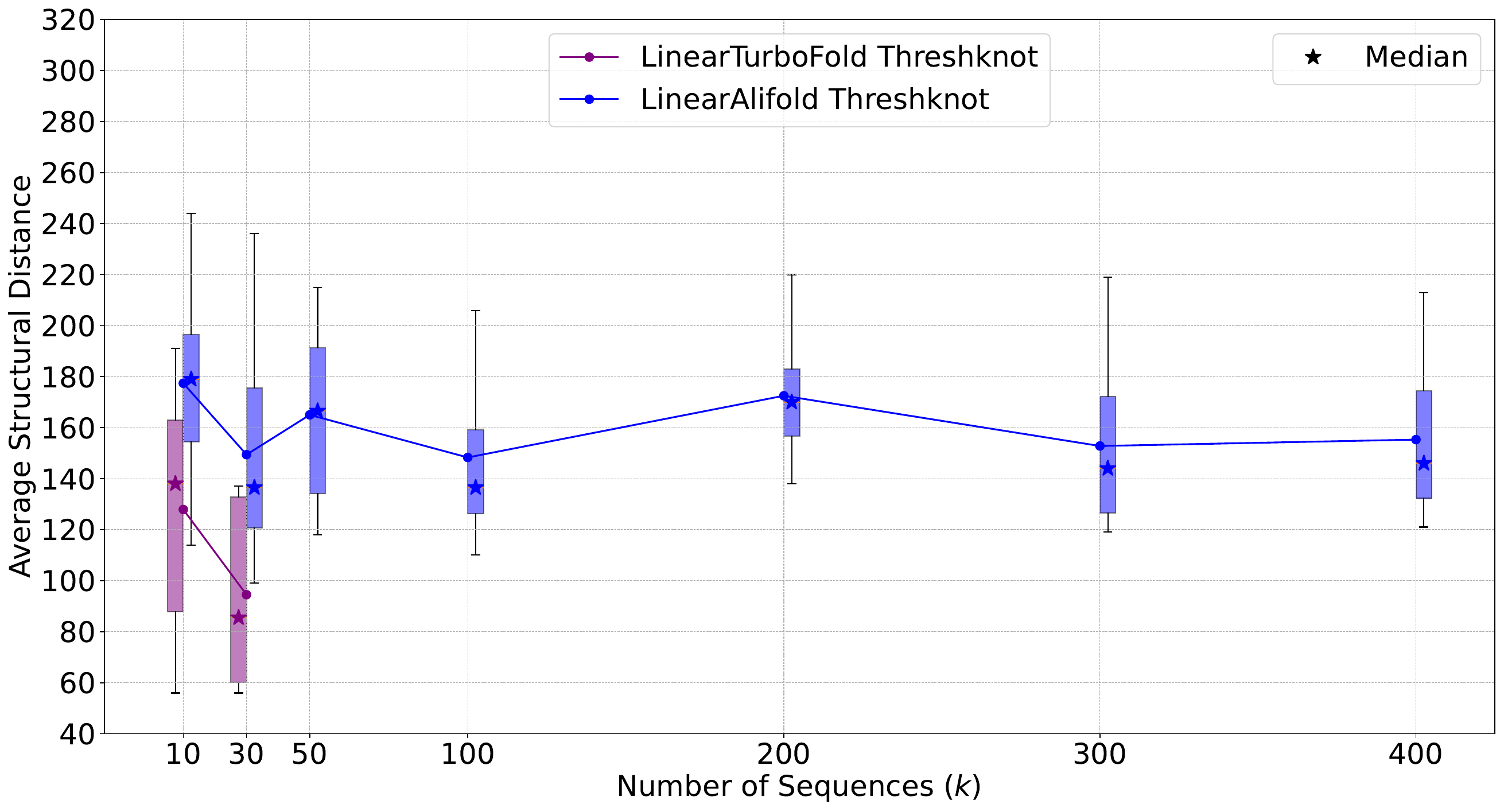} 
    & \includegraphics[width=.48\linewidth]{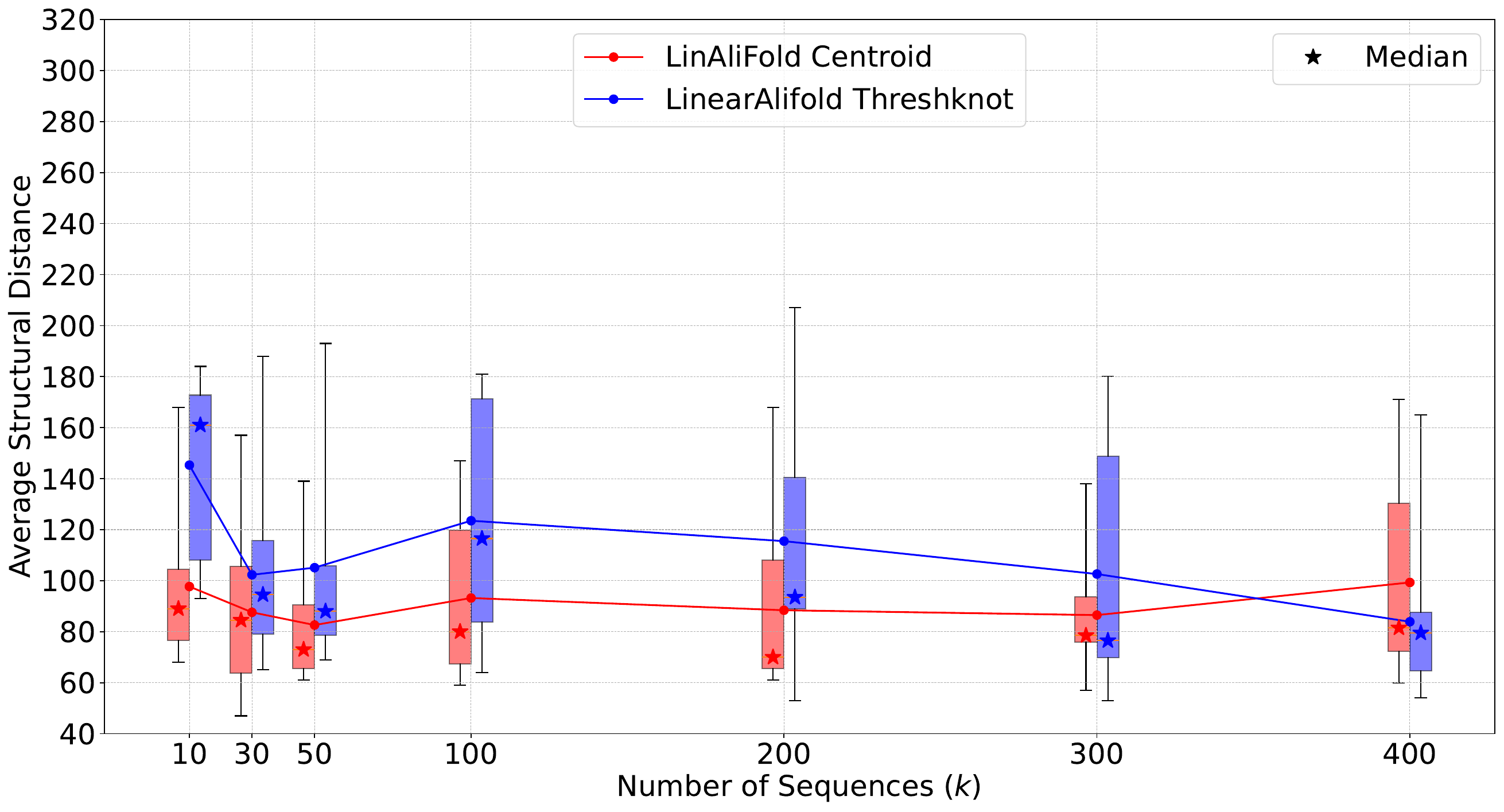} \\
    
    & \multicolumn{1}{c}{\textbf{E}} & \multicolumn{1}{c}{\textbf{F}} \\
    & \includegraphics[width=.48\linewidth]{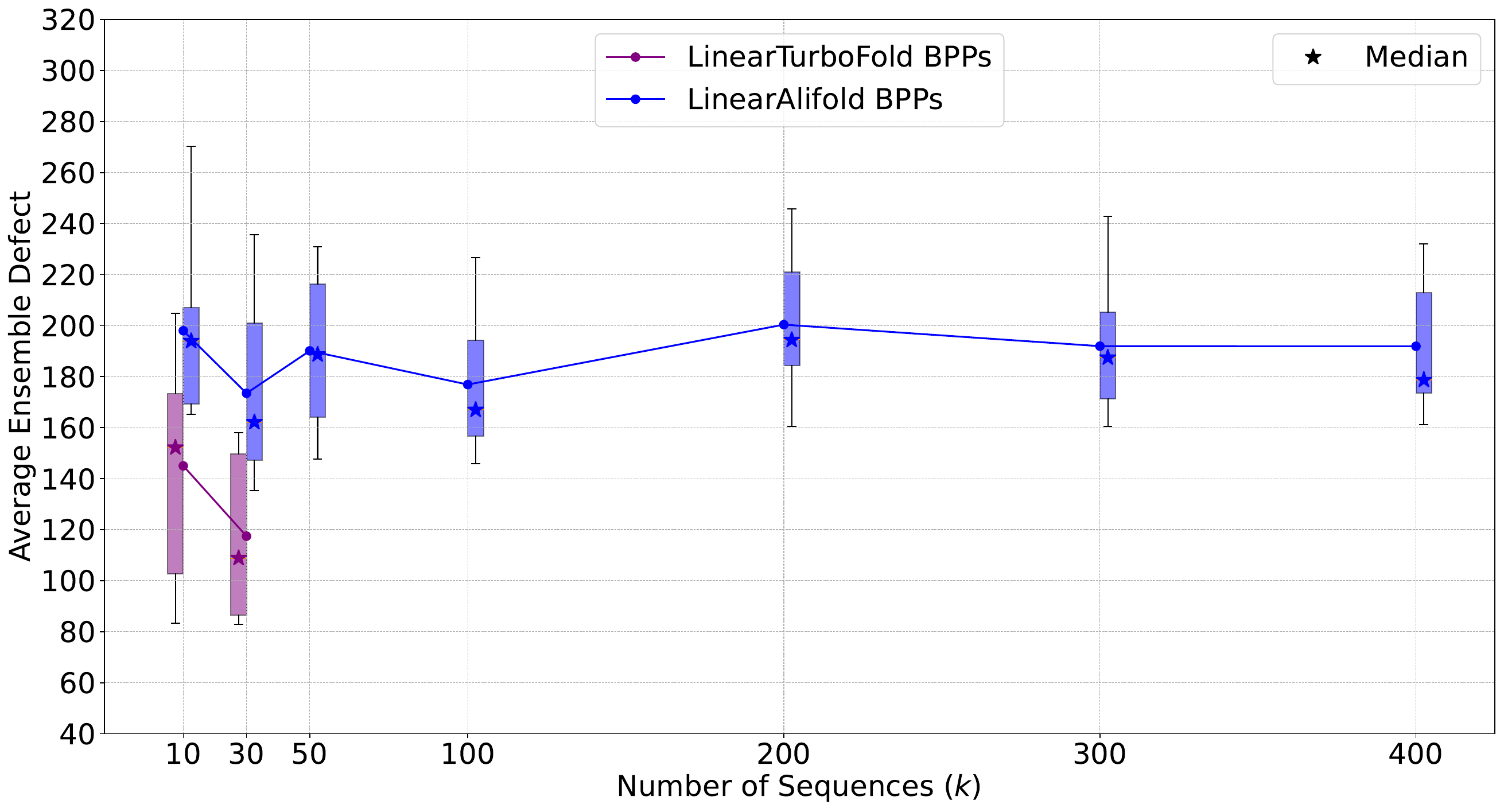} 
    & \includegraphics[width=.48\linewidth]{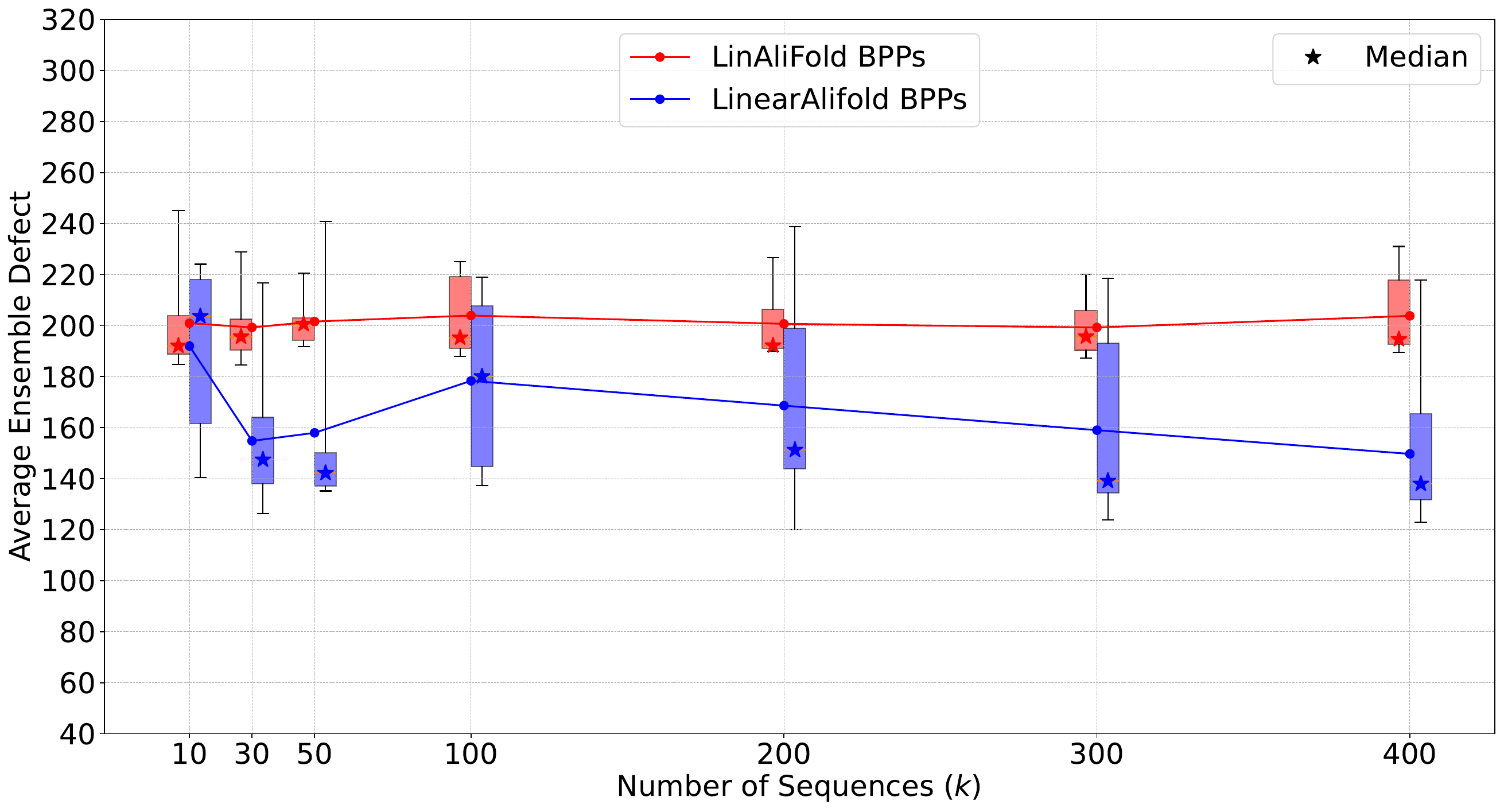} \\

    \end{tabular}

    
    \caption{Box plot of structural distance and ensemble defect (of 5' and 3' UTRs) against the number of sequences ($k$) for different energy models and methods, with the structural distance evaluated using the Huston et al.~reference structure as the reference (see Fig.~\ref{fig:covid-si} for a similar figure with the hybrid structure as reference). 
    For each $k$, we have 10 samples, so the boxes show the 25-75 percentiles, and the whiskers represent the full range of data (1-100 percentile). 
    The curves show the mean values over 10 samples for each $k$, and the stars denote the medians.
    A--B: MFE prediction. C--D: partition-based structure prediction.
    E--F: ensemble quality.}
    %
    %
    \label{fig:huston-only}
\end{figure*}

\begin{figure*}[!t]
    \centering
    \begin{tabular}{ccc}
        \small{{\bf A:} LinearAlifold Vienna Threshknot}
        &
        \small{{\bf B:} LinearAlifold BL* Threshknot}
        &
        \small{{\bf C:} LinearTurboFold Vienna Threshknot}
        \\[0.2cm]
        \includegraphics[width=0.3\textwidth]{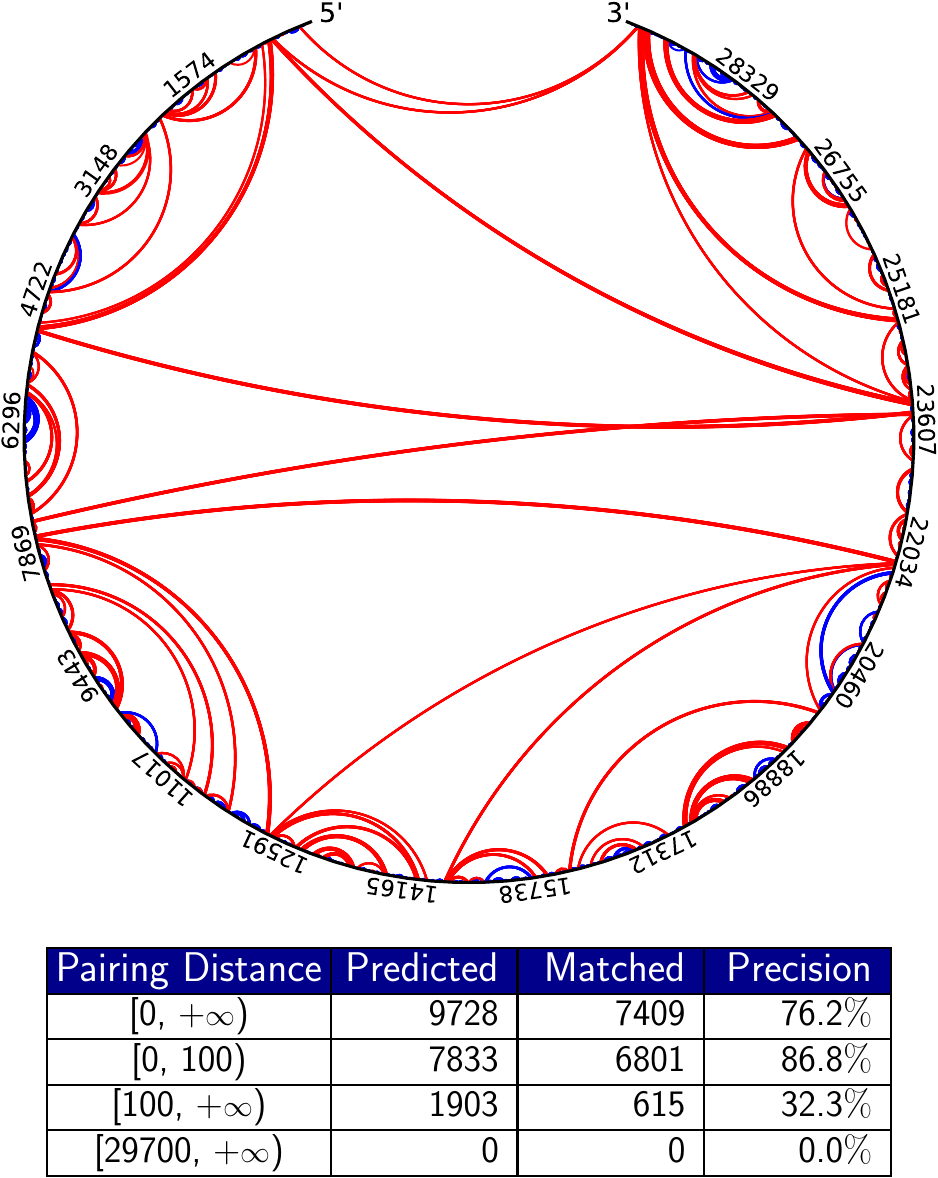}
        &
        \hspace{0.25cm}\includegraphics[width=0.3\textwidth]{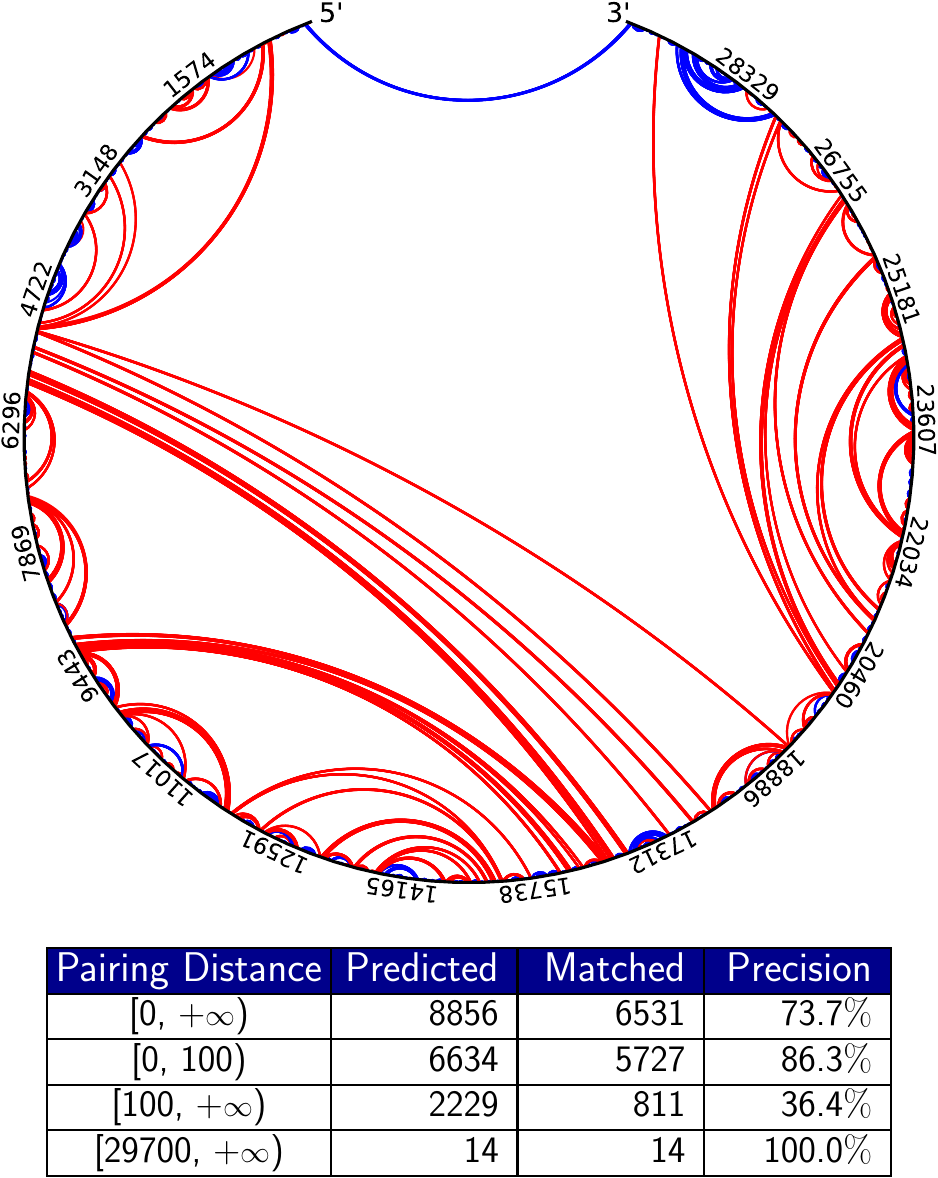}
        &
        \includegraphics[width=0.3\textwidth]{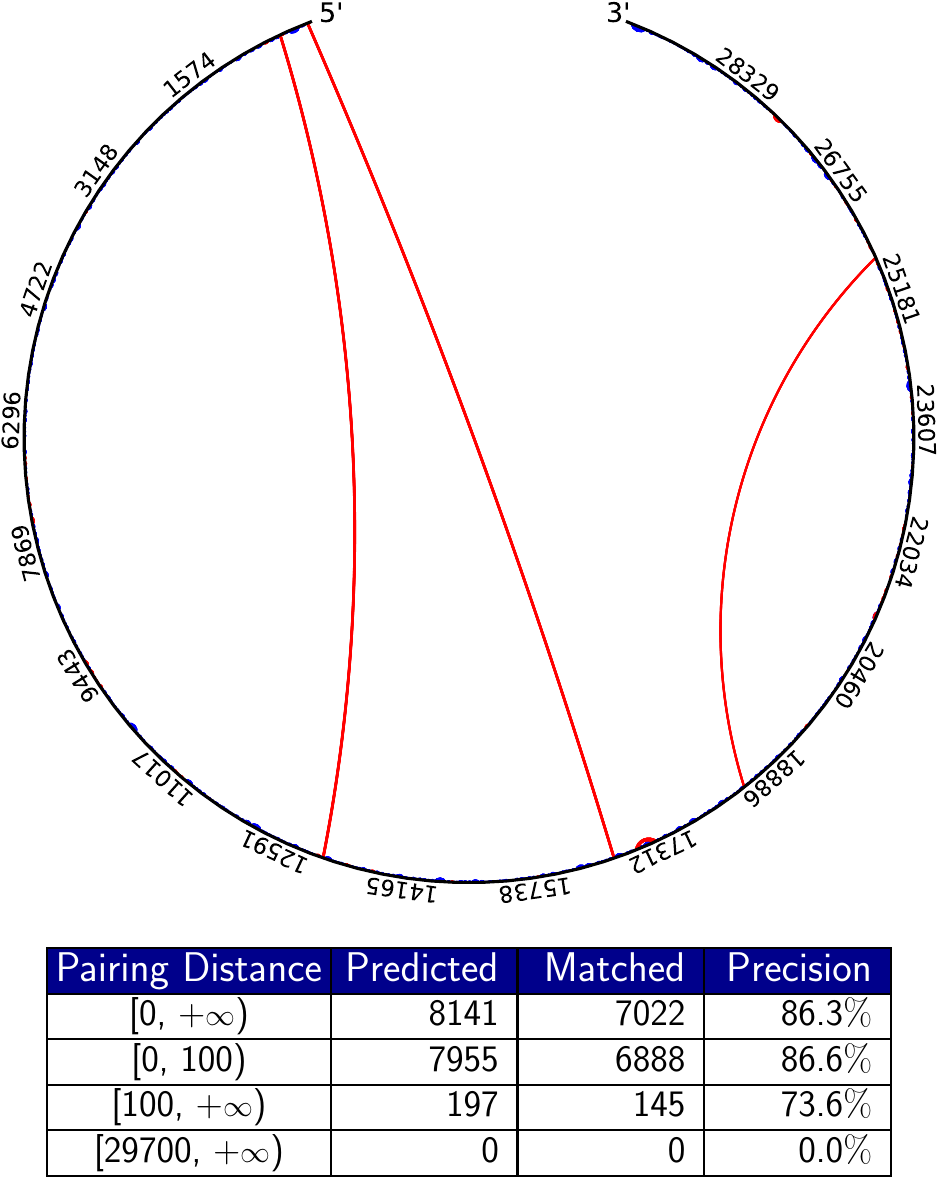}
    \end{tabular}
    \caption{COVID circular plots with Ziv et al.~range precisions for various methods. Evaluated on the COVID \( k = 30 \) sample \#5/10. Red arcs do not match any Ziv et al.~range, while blue arcs match at least one Ziv et al.~range. See also Fig.~\ref{fig:hybrid}A--C.}
    \label{fig:circular-ziv}
\end{figure*}

\end{document}